# DYNAMICS OF COSMIC FLOWS

**Avishai Dekel**

Racah Institute of Physics, The Hebrew University of Jerusalem
dekel@vms.huji.ac.il

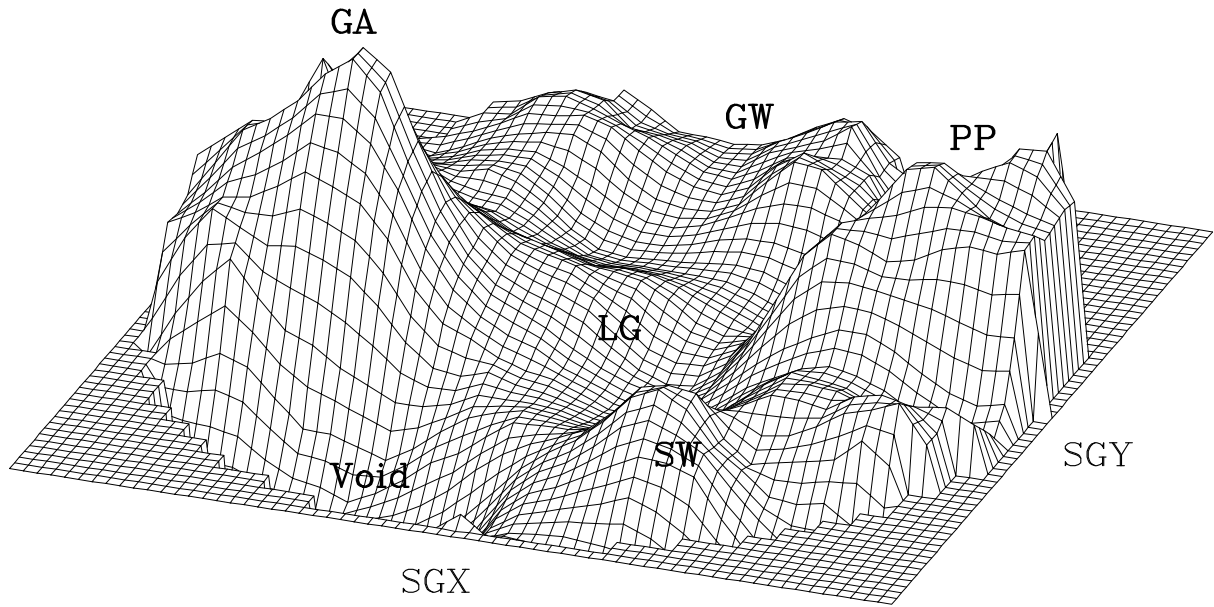

# CONTENTS

# 1. INTRODUCTION

The editors suggested a review entitled "Are There Large-Scale Motions in the Universe?". The answer is "yes", in the sense that the interpretation of the data as motions is the simplest model, so far consistent with all other available data under the current "standard model" of physical cosmology. I will review tests which could have ended up falsifying this model and failed, but the scope of this review is much extended as the field has developed far beyond the question of existence of motions. Having adopted the motions as a working hypothesis, the study of *large-scale dynamics* is becoming a mature scientific field where observation and theory are confronted in a quantitative way. It is this area of major activity in cosmology that is addressed here.

I have made no attempt to provide a complete reference list, nor have I tried a balanced discussion of all the issues of relevance and all the authors involved. My goal is to provide a critical account of some of the issues which I find important, with emphasis on theoretical implications. In many cases I am not careful in giving proper credit by quoting only a recent paper automatically implying "and references therein". The reader is referred to a comprehensive, observation-oriented review of large-scale motions in historical perspective by Burstein (1990b), a detailed review of distance indicators in a collection of essays by Jacoby *et al.* (1992), and to "Principles of Physical Cosmology" by Peebles (1993). I apologize for tending to describe in more detail work which I was involved in and therefore know better.

The current phase of the field was seeded by two major developments. One was the confirmation of the dipole moment in the Cosmic Microwave Background (CMB) (Corey & Wilkinson 1976; Smoot *et al.* 1977), indicating via Doppler shift that the Local Group of galaxies (LG) is *moving* at $\sim 600$ km s$^{-1}$ relative to the cosmological frame defined by the CMB. The other was the invention of methods for inferring distances *independent of redshifts* based on intrinsic relations between galaxy quantities (§3; Tully & Fisher 1977, TF; Faber & Jackson 1976, FJ). The radial peculiar velocity of a galaxy (the "velocity" $u$) is the difference between its total radial velocity as read from the redshift (the "redshift" $z$) and the Hubble velocity at its true distance (the "distance" $r$). Improved versions of these methods reduced the distance errors to the level of 15-21% which, with several hundred measured galaxies across the sky, enabled modeling the large-scale velocity field in terms of few-parameter "toy" models (§4.1), starting with a Virgo-centric infall (Aaronson *et al.* 1982b) and ending with spherical infall into a "Great Attractor" (GA, Lynden-Bell *et al.* 1988). The finding by the "seven samurai" (7S, Burstein *et al.* 1986) that the LG participates in a large streaming motion launched the present high-profile activity in this field. The toy modeling is gradually being replaced by non-parametric methods, where the full velocity *field* is reconstructed based on properties of gravitational flows (§4) and the associated mass-density fluctuation field is recovered from the spatial velocity derivatives (§2). With no simplified geometry imposed, the motions are not associated with single specific "sources"; the gravitational acceleration is an integral of a continuous density field consisting of swells and troughs simultaneously pulling and pushing.

A parallel major development has been of all-sky magnitude-limited redshift surveys with many thousands of galaxies, starting with the CfA and SSRS optical surveys and continuing with the very useful recent surveys based on the IRAS satellite (§5). The large-scale inhomogeneity in the galaxy distribution (*e.g.* de Lapparent *et al.* 1986) provided a clear hint for associated motions. An all-sky redshift survey can be converted into a galaxy-density field and then integrated to derive a *predicted* velocity field under the assumption of gravity and a certain "biasing" relation between galaxies and mass. The comparison of the fields obtained from redshifts to those obtained from velocities is at the heart of the research of large-scale structure (LSS), and the results carry major implications (§6.2, 8.2).

Data of both types are rapidly accumulating, and a major effort is directed at reducing the *errors* and carefully estimating those which remain, to enable quantitative testing of LSS formation theories. The standard theory consists of several *working hypotheses* which one tries to falsify by the observations or, if found consistent, to determine the characteristic model parameters. The hypotheses, which will be elaborated on later, can be listed as follows:

H1. The background *cosmology* is the standard homogeneous Friedman Robertson Walker model, possibly with an Inflation phase, where the CMB defines a cosmological "rest frame". If so, then one wishes to determine the cosmological density parameter $\Omega$ (and the cosmological constant $\Lambda$ and the Hubble constant $H$).

H2. The structure originated from a random field of small-amplitude initial density *fluctuations*. If so, the goal is to find out whether they were Gaussian, whether the power spectrum (PS) was scale-invariant (power index $n = 1$), and whether the energy density was perturbed adiabatically or in an isocurvature manner.

H3. The spectrum of fluctuations was filtered during the radiation-plasma era in a way characteristic of the nature of the *dark matter* (DM) which dominates the mass density. The DM could be baryonic or non-baryonic. If non-baryonic it could be "hot" or "cold" depending on when it became non-relativistic.

H4. The fluctuations grew by *gravitational instability* (GI) into the present LSS. This is a sufficient but not necessary condition for:

   H4a. The quasi-linear velocity field smoothed over a sufficiently-large scale is *irrotational*.

   H4b. The galaxies *trace* a unique underlying velocity field, apart from possible "velocity bias" of $\sim 10\%$ on small scales.

H5. The density fluctuations of visible galaxies are correlated with the underlying mass fluctuations. If this relation is roughly linear, then the linearized *continuity* equation in GI implies a relation between velocity and galaxy density. If so, the characteristic parameter is the *density-biasing* factor $b$.

H6. The TF and $D_n$–$\sigma$ methods measure true *distances*, which allow the reconstruction of a large-scale velocity field with known systematic biases under control.

This review is geared toward the confrontation of observations with these hypotheses. The relevant observations can be classified into the following three major categories:

O1. Angular fluctuations in the CMB temperature at various angular scales.

O2. The distribution of luminous objects on the sky and in redshift space.

O3. Peculiar velocities of galaxies along the line of sight.

Note that O1 and O3 are related to dynamical ingredients H1-4 and H6, bypassing the uncertain nature of galaxy-density biasing H5. Also, O1 and O2 refer to the theory independently of H4a,b and H6, which address the velocities and their analysis.

## 2. GRAVITATIONAL INSTABILITY

This section provides a brief account of the standard theory of GI, and of the linear and quasi-linear approximations which serve the analysis of motions. Let $\boldsymbol{x}, \boldsymbol{v}$, and $\Phi_g$ be the position, peculiar velocity and peculiar gravitational potential in comoving distance units, corresponding to $a\boldsymbol{x}$, $a\boldsymbol{v}$, and $a^2 \Phi_g$ in physical units, with $a(t)$ the universal expansion factor. Let the mass-density fluctuation be $\delta \equiv (\rho - \bar\rho)/\rho$. The equations governing the evolution of fluctuations of a pressureless gravitating fluid in a standard cosmological background during the matter era are the *Continuity* equation, the *Euler* equation of motion, and the *Poisson* field equation (*e.g.* Peebles 1993):

$$\dot\delta + \boldsymbol{\nabla} \cdot \boldsymbol{v} + \boldsymbol{\nabla} \cdot (\boldsymbol{v}\delta) = 0 , \tag{1}$$

$$\dot{\boldsymbol{v}} + 2H\boldsymbol{v} + (\boldsymbol{v} \cdot \boldsymbol{\nabla})\boldsymbol{v} = -\boldsymbol{\nabla}\Phi_g , \tag{2}$$

$$\boldsymbol{\nabla}^2 \Phi_g = (3/2)H^2 \Omega \delta , \tag{3}$$

with $H$ and $\Omega$ varying in time. The dynamics do not depend on the value of the Hubble constant H; it is set to unity by measuring distances in $\mathrm{km\,s^{-1}}$ (1 $h^{-1}$Mpc = 100 $\mathrm{km\,s^{-1}}$).

In the *linear* approximation, the GI equations can be combined into a time evolution equation, $\ddot\delta + 2H\dot\delta = (3/2)H^2\Omega\delta$. The growing mode of the solution, $D(t)$, is irrotational and can be expressed in terms of $f(\Omega) \equiv H^{-1}\dot D/D \approx \Omega^{0.6}$ (see Peebles 1993, eq. 5.120). The linear relation between density and velocity is

$$\delta = \delta_0 \equiv -(Hf)^{-1}\boldsymbol{\nabla} \cdot \boldsymbol{v} . \tag{4}$$

The use of $\delta_0$ is limited to the small dynamical range between a *few* tens of megaparsecs and the $\sim 100\ h^{-1}$Mpc extent of the current samples. In contrast the sampling of galaxies enables reliable dynamical analysis with smoothing as small as $\sim 10\ h^{-1}$Mpc, where $|\boldsymbol{\nabla}\cdot\boldsymbol{v}|$ obtains values $\geq 1$ so quasi-linear effects play a role. Unlike the strong non-linear effects in virialized systems which erase any memory of the initial conditions, mild non-linear effects carry crucial information about the formation of LSS, and should therefore be treated carefully. Figure 1 shows that $\delta_0$ becomes a severe underestimate at large $|\delta|$. This explains why eq. (4) is invalid in the non-linear epoch even where $\delta = 0$; the requirements that $\int \delta\, d^3x = 0$ by definition and $\int \boldsymbol{\nabla}\cdot\boldsymbol{v}\, d^3x = 0$ by isotropy imply $-\boldsymbol{\nabla}\cdot\boldsymbol{v} > \delta$ at $|\delta| \ll 1$. Fortunately, the small variance of $\boldsymbol{\nabla}\cdot\boldsymbol{v}$ given $\delta$ promises that some function of the velocity derivatives may be a good local approximation to $\delta$.

A basis for useful *quasi-linear* relations is provided by the *Zel'dovich* (1970) approximation (Z). The displacements of particles from their initial, Lagrangian positions $\boldsymbol{q}$ to their Eulerian positions $\boldsymbol{x}$ at time $t$ are assumed to have a universal time dependence,

$$\boldsymbol{x}(\boldsymbol{q}, t) - \boldsymbol{q} = D(t)\boldsymbol{\psi}(\boldsymbol{q}) = f^{-1}\boldsymbol{v}(\boldsymbol{q}, t) . \tag{5}$$

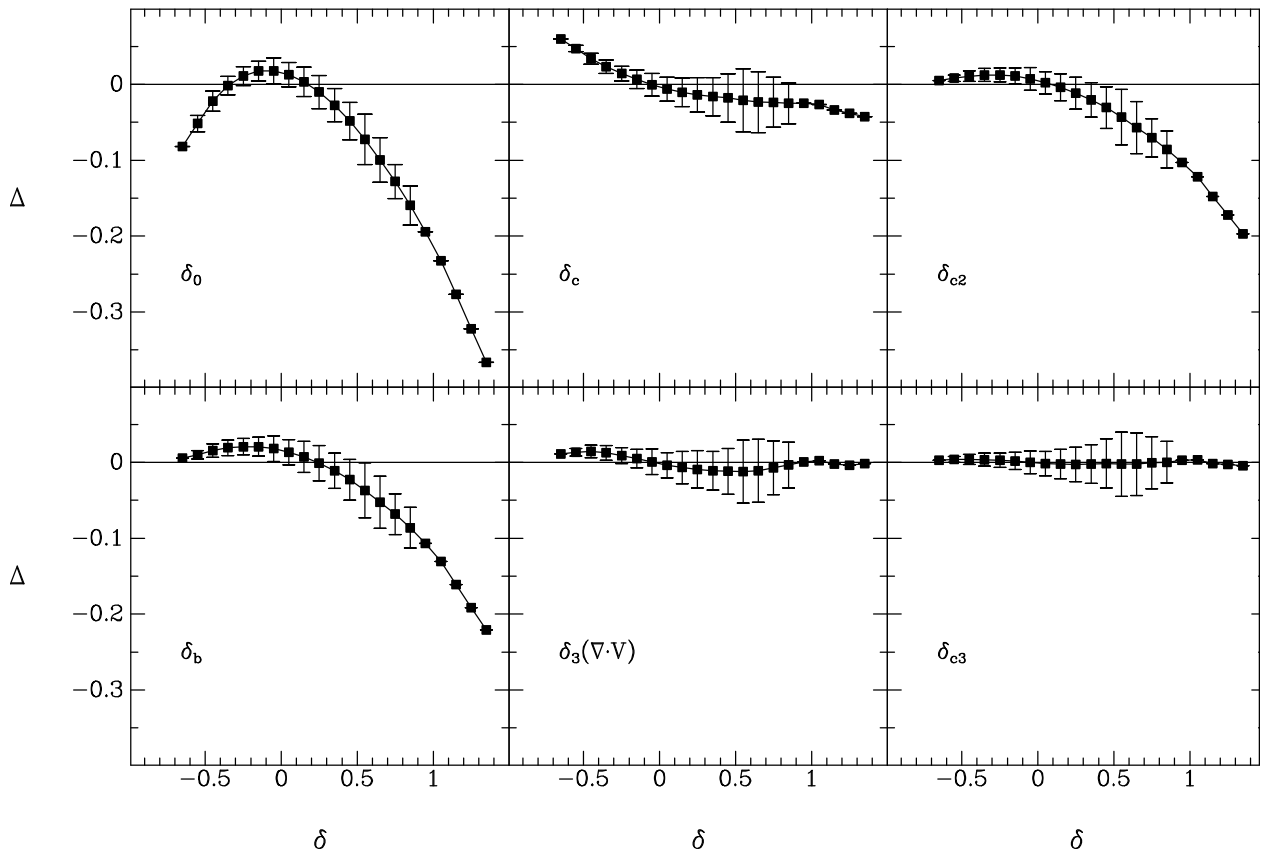

**Figure 1:** Quasi-linear velocity-to-density approximations. $\Delta \equiv \delta_{approx}(v) - \delta_{true}$. The mean and standard deviation are from large standard-CDM N-body simulations normalized to $\sigma_8 = 1$, Gaussian-smoothed with radius 12 h$^{-1}$Mpc (see Mancinelli et al. 1994). Note the factor of 5 difference in scale between the axes.

For the purpose of approximating GI, the Lagrangian Z approximation can be interpreted in *Eulerian* space,

$$q(x) = x - f^{-1}v(x) , \qquad (6)$$

provided that the flow is *laminar* (*i.e.* that multi-streams are appropriately smoothed over). The solution of the continuity equation then yields (Nusser et al. 1991)

$$\delta_c(x) = \|I - f^{-1}\partial v/\partial x\| - 1 , \qquad (7)$$

where the bars denote the Jacobian determinant, and $I$ is the unit matrix. The Z displacement is first order in $f^{-1}$ and $v$, so $\delta_c$ involves second- and third-order terms ($m_{v2}, m_{v3}$) as well. The relation (7) is not easily invertible to provide $\nabla \cdot v$ or $v$ when $\delta$ is given, but a useful approximation derived from simulations is

$$\nabla \cdot v = -f\delta/(1 + 0.18\delta) . \qquad (8)$$

A modified approximation, which is derived by adding a second-order term to the Z displacement (Moutarde et al. 1991) and truncating all the expressions at second order while solving the continuity equation is (Gramann 1993a)

$$\delta_{c2} = -f^{-1}\boldsymbol{\nabla}\cdot\boldsymbol{v} + (4/7)f^{-2}m_{v2} \ , \quad m_{v2} \equiv \sum_i \sum_{j>i}(\partial_i v_i \partial_j v_j - \partial_j v_i \partial_i v_j) \ . \qquad (9)$$

The factor 4/7 replaces 1 in the second-order term of $\delta_c$. While terms are kept to second-order, it is still not an exact solution to the second-order equations of GI. This relation can be inverted in second-order to provide $\boldsymbol{\nabla}\cdot\boldsymbol{v}$ given $\delta$, with $m_{v2}$ replaced by an analogous expression $m_{g2}$, involving the gravitational acceleration $\boldsymbol{g}$.

Since the variance of $\delta$ given $\boldsymbol{\nabla}\cdot\boldsymbol{v}$ is small, one expects that a non-linear function of $\boldsymbol{\nabla}\cdot\boldsymbol{v}$ which properly corrects for the systematic deviation can be a good quasi-linear approximation to $\delta$ (and vice versa). Assuming Gaussian initial fluctuations, Bernardeau (1992) found a solution in the limit of vanishing variance:

$$\delta_b = [1 - (2/3)f^{-1}\boldsymbol{\nabla}\cdot\boldsymbol{v}]^{3/2} - 1 \ , \qquad (10)$$

which is easily invertible. A polynomial expansion with non-vanishing variance should have the form (Zehavi & Dekel, in prep.)

$$\delta_n(\boldsymbol{\nabla}\cdot\boldsymbol{v}) = -f^{-1}\boldsymbol{\nabla}\cdot\boldsymbol{v} + a_2 f^{-2}[(\boldsymbol{\nabla}\cdot\boldsymbol{v})^2 - \mu_2] + a_3 f^{-3}[(\boldsymbol{\nabla}\cdot\boldsymbol{v})^3 - \mu_3] + \dots \ . \qquad (11)$$

Because the first two terms vanish when integrated over a large volume, the moments $\mu_n \equiv \langle(\boldsymbol{\nabla}\cdot\boldsymbol{v})^n\rangle$ must be subtracted off to make the $n^{th}$-order term vanish as well. The coefficients can be crudely approximated analytically (*e.g.* Bernardeau 1992) or, using CDM simulations and Gaussian smoothing, the best coefficients are $a_2 \approx 0.3$ and $a_3 \approx -0.1$, tested for $\Omega$ values $0.1-1$ and smoothing radii $5-12$ h$^{-1}$Mpc at $\sigma_8 = 1$ ($\sigma_8$ is the *rms* of unsmoothed mass-$\delta$ in top-hat spheres of radius 8 h$^{-1}$Mpc). The structure of eq. (11) makes it *robust* to uncertain features such as $\Omega$, the shape of the fluctuation power spectrum, and the degree of non-linearity as determined by the fluctuation amplitude and the smoothing. Such robustness is crucial when using a quasilinear approximation for determining $\Omega$, for example (§8).

Fig. 1 demonstrates the accuracy of the explicit quasi-linear approximations using CDM N-body simulations and 12 h$^{-1}$Mpc smoothing (Mancinelli *et al.* 1994). $\delta_c$, of scatter $\sim 0.1$, is an excellent approximation for $\delta \geq 1$ but it is a slight overestimate at the negative tail. $\delta_{c2}$ and $\delta_b$ do better at the negative tail, but they are severe underestimates in the positive tail. $\delta_3(\boldsymbol{\nabla}\cdot\boldsymbol{v})$ is an excellent robust fit over the whole quasi-linear regime. $\delta_{c3}$ is constructed from the three terms in the expansion of $\delta_c$ in powers of $f^{-1}$ but with the numerical coefficients adjusted to achieve best fit in the simulation ($-1.05, 0.9, 1.5$ replacing unity, independent of $\Omega$).

# 3. MEASURING PECULIAR VELOCITIES

## 3.1. Distance Indicators

Measuring redshift-independent distances to many galaxies at large distances is the key to large-scale dynamics (review: Jacoby *et al.* 1992). The simplest method assumes that a certain class of objects is a "standard candle", in the sense that a distance-dependent observable is distributed intrinsically at random with small variance about a universal mean. The luminosity of an object ($\propto r^{-2}$) or its diameter (($\propto r^{-1}$), can serve as this quantity. In a pioneering study, Rubin *et al.* (1976a,b) used the brightness of giant Sc spirals to discover a net motion for the shell at 35–60 $h^{-1}$Mpc that agrees within the errors with more modern results, but the large uncertainties in this simple distance indicator made this result controversial at the time.

So far, the most useful distance indicators for LSS have been of the TF-kind, based on intrinsic relations between *two* quantities: a distance-dependent quantity such as the flux $\propto L/r^2$, and a distance-independent quantity $\sigma$ – the maximum rotation velocity of spirals (TF) or the velocity dispersion in ellipticals (FJ). The intrinsic relations are power laws, $L \propto \sigma^\beta$, *i.e.*

$$M(\eta) = a - b\eta, \qquad (12)$$

where $M \equiv -2.5\log L + const$ is the absolute magnitude and $\eta \equiv \log\sigma$. The slope $b$ can be determined empirically in clusters, where all the galaxies are assumed to be at the same distance, typically yielding $\beta \approx 3-4$, depending on the luminosity band (*e.g.* $\beta_I \approx 3$, $\beta_H \approx 4$). Then, for any other galaxy with observed $\eta$ and apparent magnitude $m \equiv -2.5\log(L/r^2) + const$, one can determine a *relative* distance via $5\log r = m - M(\eta)$. There exists a fundamental freedom in determining the *zero point*, $a$, which fixes the distances at absolute values (in km s$^{-1}$, not to be confused with $H$ which translates to Mpc). Changing $a$, *i.e.* multiplying the distances by a factor $(1+\epsilon)$ while the redshifts are fixed, is equivalent to adding a monopole Hubble-like component $-\epsilon r$ to $\boldsymbol{v}$, and an offset $3\epsilon$ to $\delta$ (eq. 4). It has been arbitrarily determined in several data sets, *e.g.* by assuming $u=0$ for the Coma cluster, but $a$ is better determined by minimizing the variance of the recovered peculiar velocity field in a large "fair" volume. The original TF technique has been improved by moving from blue to near-infrared photometry (H band, Aaronson *et al.* 1979) and recently to CCD R and I bands, where spiral galaxies are more transparent and therefore the intrinsic scatter is reduced to $\sigma_m \sim 0.33$ mag, corresponding to a relative distance error of $\Delta = (\ln 10/5)\sigma_m \approx 0.15$.

A distance indicator of similar quality for ellipticals has proved harder to achieve. Minimum variance, corresponding to $\Delta = 0.21$, was found for a revised FJ relation involving three physical quantities: $DI^\alpha \propto \sigma^\beta$ with $D$ the diameter and $I \propto L/D^2$ the surface brightness (Dressler *et al.* 1987; Djorgovski & Davis 1987). The parameters were found to be $\alpha \approx 5/6$ and $\beta \approx 4/3$. By defining from the photometry a "diameter" at a fixed value of enclosed $I$, termed $D_n$, the relation returns to a simple form, $D_n \propto \sigma^\beta$, similar to FJ but with reduced variance.

The physical origin of the scaling relations is not fully understood, reflecting our limited understanding of galaxy formation. What matters for the purpose of distance measurements is the mean empirical relation and its variance. However, one can point at an important physical difference between the two relations (Gunn 1989), which is relevant to the testing for environmental effects (§6.3). The $D_n$–$\sigma$ relation is naturally explained by virial equilibrium, $\sigma^2 \propto M/D$, and a smoothly varying $M/L \propto M^\gamma$, which together yield $DI^{1/(1+\gamma)} \propto \sigma^{2(1-\gamma)/(1+\gamma)}$, but the TF relation, involving only two of the three quantities entering the virial theorem, is more demanding – it requires an additional constraint which is probably imposed at galaxy formation.

There is some hope for reducing the error in the TF method to the $\sim 10\%$ ball park by certain modifications, *e.g.* by restricting attention to galaxies of normal morphology (Raychaudhury 1994). But the most accurate to date is the estimator based on surface-brightness fluctuations (SBF) in ellipticals (Tonry 1991), where the standard candle is the luminosity function of bright stars in the old population. These stars show up as distance-dependent fluctuations in sensitive surface-brightness measurements. The technique is being applied successfully out to $\sim 30$ h$^{-1}$Mpc (*e.g.* Dressler 1994), with the improved accuracy of $\sim 5\%$ enabling high-resolution non-linear analysis, and it can be of great value for LSS if applied at larger distances. The need to remove sources of unwanted fluctuations such as globular clusters requires high resolution observations which could be achieved by HST or adaptive optics.

The prospects for the future can be evaluated by estimating the length scale over which LSS dynamics can be studied using a distance indicator of relative error $\Delta$. The error in a velocity derived from $N$ galaxies at a distance $\sim r$ is $\sigma_V \sim r\Delta/\sqrt{N}$. Let the mean sampling density be $\bar{n}$. Let the desired quantity be the mean velocity $V$ in spheres of radius $R$, and assume that its true *rms* value is $V_{20}$ at $R = 20$ h$^{-1}$Mpc and $\propto R^{-(n+1)}$ on larger scales, with $n$ the effective power index of the fluctuation spectrum near $R$. Then the relative error in $V$ is

$$\frac{\sigma_V}{V} \approx 0.033 \left(\frac{\bar{n}}{0.01}\right)^{-1/2} \left(\frac{\Delta}{0.15}\right) \left(\frac{V_{20}}{500}\right)^{-1} \left(\frac{R}{20}\right)^{n+1/2} \frac{r}{R} , \qquad (13)$$

where distances are measured in h$^{-1}$Mpc. The observations indicate that $V_{20} \sim 500$ km s$^{-1}$ and $n \sim -0.5$ for $R = 20 - 60$ h$^{-1}$Mpc (§7.1). Thus, with ideal sampling of $\bar{n} \sim 0.01$ (h$^{-1}$Mpc)$^{-3}$, the relative error is always only a few percent of $r/R$. This means that LSS motions can in principle be meaningfully studied at all distances $r$ with smoothing $R \sim 0.1r$, as long as $n \sim -0.5$ at the desired $R$. Since $n$ seems to be negative out to $\sim 100$ h$^{-1}$Mpc (§7.1), dense deep TF samples are potentially useful out to several hundred megaparsecs. However, several technical difficulties pose a serious challenge at such distances. For example, the calibration requires faint cluster galaxies which are harder to identify, aperture effects become severe, the spectroscopy capability is limited.

### 3.2. Malmquist Biases

The random scatter in the distance estimator is a source of severe systematic biases in the inferred distances and peculiar velocities, which are generally termed "Malmquist" biases but should carefully be distinguished from each other (*e.g.* Lynden-Bell *et al.* 1988; Willick 1994a,b).

The calibration of the TF relation is affected by the *selection bias* (or *calibration bias*). A magnitude limit in the selection of the sample used for calibration at a fixed *true* distance (*e.g.* in a cluster) tilts the "forward" TF regression line of $M$ on $\eta$ towards bright $M$ at small $\eta$ values. The bias extends to all values of $\eta$ when objects at a large range of distances are used for the calibration. This bias is inevitable when the dependent quantity is explicitly involved in the selection process, and it occurs to a certain extent even in the "inverse" relation $\eta(M)$ due to existing dependences of the selection on $\eta$. Fortunately, the selection bias can be corrected once the selection function is known (*e.g.* Willick 1991; 1994a).

The TF inferred distance, $d$, and the mean peculiar velocity at a given $d$, suffer from an *inferred-distance bias*, which we term hereafter "M" bias. I comment later (§4.4) on a possible way to avoid the M bias by performing an inverse analysis in $z$-space, at the expense of a more complicated procedure and other biases. Here I focus on a statistical way for correcting the M bias within the simpler forward TF procedure in $d$-space. This bias can also be corrected in an inverse TF analysis in $d$-space, using the selection function $S(d)$ which is in principle derivable from the sample itself (Landy & Szalay 1992).

The current POTENT procedure uses the forward TF relation in $d$-space. If $M$ is distributed normally for a given $\eta$, with standard deviation $\sigma_m$, then the TF-inferred distance $d$ of a galaxy at a true distance $r$ is distributed log-normally about $r$, with relative error $\Delta \approx 0.46\sigma_m$. Given $d$, the expectation value of $r$ is (*e.g.* Willick 1991):

$$E(r|d) = \frac{\int_0^\infty r P(r|d) \mathrm{d}r}{\int_0^\infty P(r|d) \mathrm{d}r} = \frac{\int_0^\infty r^3 n(r) \exp\left(-\frac{[\ln(r/d)]^2}{2\Delta^2}\right) \mathrm{d}r}{\int_0^\infty r^2 n(r) \exp\left(-\frac{[\ln(r/d)]^2}{2\Delta^2}\right) \mathrm{d}r} , \qquad (14)$$

where $n(r)$ is the number density in the underlying distribution from which galaxies were selected (by quantities that do not explicitly depend on $r$). The deviation of $E(r|d)$ from $d$ reflects the bias. The homogeneous part (HM) arises from the geometry of space – the inferred distance $d$ underestimates $r$ because it is more likely to have been scattered by errors from $r > d$ than from $r < d$, the volume being $\propto r^2$. If $n = const$, equation (10) reduces to

$$E(r|d) = d\, e^{3.5\Delta^2} , \qquad (15)$$

in which the inferred distances are simply multiplied by a factor, 8% for $\Delta = 0.15$, equivalent to changing the zero-point of the TF relation. The HM bias is corrected this way on a regular basis since 7S.

Fluctuations in $n(r)$ are responsible for the inhomogeneous bias (IM), which is worse because it systematically enhances the inferred density perturbations and the value of $\Omega$ inferred from them. If $n(r)$ is varying slowly with $r$, and if $\Delta \ll 1$, then eq. (14) reduces to

$$E(r|d) = d \left[1 + 3.5\Delta^2 + \Delta^2 \left(\frac{d\ln n}{d\ln r}\right)_{r=d}\right] , \qquad (16)$$

showing the dependence on $\Delta$ and the gradients of $n(r)$. To illustrate, consider a lump of galaxies at one point $r$ with $u = 0$. Their inferred distances are randomly scattered to the foreground and background of $r$. With all galaxies having the same $z$, the inferred $u$ on either side of $r$ mimic a spurious infall towards $r$, which is interpreted dynamically as a spurious overdensity at $r$.

In the current data for POTENT analysis (§4.2) IM bias is corrected in two steps. First, the galaxies are heavily grouped in $z$-space (Willick *et al.* 1994), reducing the distance error of each group of $N$ members to $\Delta/\sqrt{N}$ and thus significantly weakening the bias. Then, the noisy inferred distance of each object, $d$, is replaced by $E(r|d)$ (eq. 14), with an assumed $n(r)$ properly corrected for grouping. This procedure has been tested using realistic mock data from N-body simulations (Kolatt *et al.*, in prep.), showing that IM bias can be reduced to a few percent. The practical uncertainty is in $n(r)$, which can be approximated by the high-resolution density field of IRAS or optical galaxies (§5), or by the recovered mass-density itself in an iterative procedure under some assumption about how galaxies trace mass. The second-step correction to $\delta$ recovered by POTENT is $<20\%$ even at the highest peaks (Dekel *et al.* 1994).

### 3.3. Homogenized Catalogs

Several samples of galaxies with TF or $D_n$–$\sigma$ measurements have accumulated in the last decade. Assuming that all galaxies trace the same underlying velocity field (§6.3), the analysis of large-scale motions greatly benefits from merging the different samples into one self-consistent catalog. The observers differ in their selection procedure, the quantities they measure, the method of measurement and the TF calibration techniques, which cause systematic errors and make the merger non-trivial. The original merged set, compiled by D. Burstein (Mark II) and used in the first application of POTENT (Bertschinger *et al.* 1990), consisted of 544 ellipticals and S0's (Lynden-Bell *et al.* 1988; Faber *et al.* 1989; Lucey & Carter 1988; Dressler & Faber 1991) and 429 spirals (Aaronson *et al.* 1982a; Aaronson *et al.* 1986; 1989; Bothun *et al.* 1984). The current merged set (Willick *et al.* 1994, Mark III of the Burstein series) consists of $\sim 2850$ spirals (Mark II plus Han & Mould 1990, 1992; Mould *et al.* 1991; Willick 1991; Courteau 1992; Mathewson *et al.* 1992) and the ellipticals of Mark II. This sample enables a reasonable recovery of the dynamical fields with $\sim 12$ h$^{-1}$Mpc smoothing in a sphere of radius $\sim 60$ h$^{-1}$Mpc about the LG, extending to $\sim 80$ h$^{-1}$Mpc in certain regions (§4). Part of the data are shown in Figure 2.

As carried out by this group, merger of catalogs involves the following major steps: (1) Standardizing the selection criteria, *e.g.* rejecting galaxies of high inclination or low $\eta$ which are suspected of large errors and sharpening any $z$ cutoff. (2) Rederiving a provisional TF calibration for each data set using Willick's algorithm (1994a) which simultaneously groups, fits and corrects for selection bias, and then verifying that inverse-TF distances to clusters are similar to the forward-TF distances. (3) Starting with one data set, adding each new set in succession using the galaxies in common to adjust the TF parameters of the new set if necessary. (4) Using only one measurement per galaxy even if it was observed by more than one observer to ensure well defined errors, and using multiple observations for a "cluster" only if the overlap is small (*e.g.* $<50\%$). (5) Adding the ellipticals from Mark II, allowing for a slight zero-point shift (§6.3). Such a careful calibration and merger procedure is *crucial* for reliable results – in several cases it produced TF distances substantially different from those quoted by the original authors.

### 4. ANALYSIS OF OBSERVED PECULIAR VELOCITIES

Given radial peculiar velocities $u_i$ sparsely sampled at positions $\boldsymbol{x}_i$ over a large volume, with random errors $\sigma_i$, the first goal is to extract the underlying three-dimensional velocity field, $\boldsymbol{v}(\boldsymbol{x})$. Under GI, this velocity field is subject to certain constraints, *e.g.* being associated with a mass-density fluctuation field, $\delta(\boldsymbol{x})$ – the other target for recovery. A field can be defined by a parametric model or by the field values at grid points. The number of independent parameters or grid points should be much smaller than the number of data points in view of the noisy data.

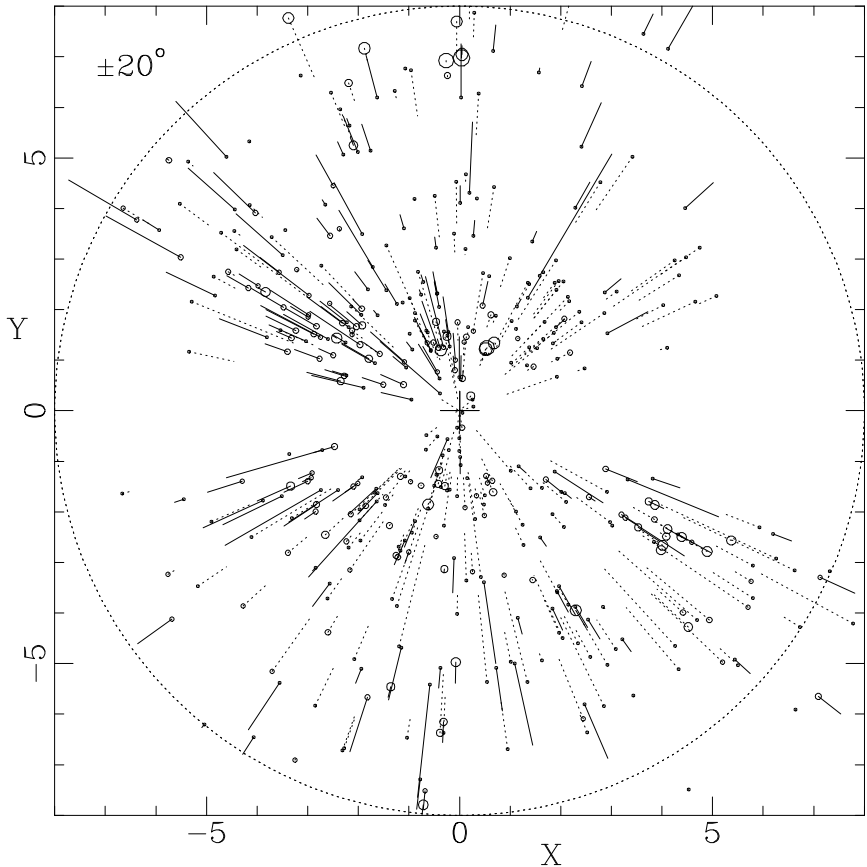

**Figure 2:** Inferred radial peculiar velocities of grouped galaxies in a $\pm 20°$ slice about the Supergalactic plane from the homogenized Mark III catalog (Willick *et al.* 1994). Distances and velocities are in 1000 km s$^{-1}$. The area of each circle marking the object position is proportional to the object richness. This slice contains 453 objects made of 1124 galaxies out of 1214 objects in the whole volume. Solid and dashed lines distinguish between outgoing and incoming objects. The positions and velocities are corrected for IM bias. Note the GA convergence (left) and the PP convergence (right-bottom).

### 4.1. Toy Models

Given a model velocity field $\boldsymbol{v}(\alpha_k, \boldsymbol{x})$, the free parameters $\alpha_k$ can be determined globally by minimizing a weighted sum of residuals, *e.g.*

$$-\log L \propto \sum_i W_i \left[ u_i - \hat{\boldsymbol{x}}_i \cdot \boldsymbol{v}(\alpha_k, \boldsymbol{x}_i) \right]^2 . \qquad (17)$$

If the errors are Gaussian and $W_i \propto \sigma_i^2$ then $L$ is the likelihood, and it is a useful approximation for log-normal errors as well. The model could first be a few-parameter

"toy" model with simple geometry. Already the simplest bulk-flow model, $\boldsymbol{v}(\boldsymbol{x}) = \boldsymbol{B}$ corresponding to $\delta = 0$, is of interest because the data clearly show a bulk flow component of several hundred km s$^{-1}$ in our neighborhood (§7.1). Another simple model is of spherical symmetry, expected to be a reasonable fit in voids and in regions dominated by one high density peak (Bardeen *et al.* 1986). The velocity profile as a function of distance $r$ from the infall center is not particularly constrained by GI, and a specific profile which proved successful is

$$v(r) = -v_{lg} \left(\frac{r}{r_{lg}}\right) \left[\frac{(r_{lg}^2 + r_c^2)}{(r^2 + r_c^2)}\right]^{(n+1)/2}. \tag{18}$$

The center is specified by its angular position and its distance $r_{lg}$ from the LG, and the profile is characterized by its value $v_{lg}$ at the LG, a core radius $r_c$, and a power index $n$. For $r \ll r_c$ the velocity rises $\propto r$ and for $r \gg r_c$ it falls off $\propto r^{-n}$. The associated density profile is given in the linear approximation by the divergence $\delta = -f^{-1} r^{-1} \partial_r [r\, v(r)]$. The 7S ellipticals were modeled by a spherical infall model termed "The Great Attractor" (GA), with $r_{lg} = 42$ h$^{-1}$Mpc toward $(l,b) = (309°, +18°)$, $v_{lg} = 535$ km s$^{-1}$, $r_c = 14.3$ h$^{-1}$Mpc, and $n = 1.7$ (Faber & Burstein 1988). A similar model vaguely fits the local infall of spirals into Virgo (Aaronson *et al.* 1982b) The merged Mark II data is well fitted by a multi-parameter hybrid consisting of a GA infall, a Virgo-centric infall, and a "local anomaly" – a bulk flow shared by the $\sim 10$ h$^{-1}$Mpc-local neighborhood of 360 km s$^{-1}$ perpendicular to the supergalactic plane.

The toy models provide an intuitive picturing of the large-scale motions with clues about the associated mass sources, and they can be used as simple statistics for the comparison with theory (e.g. the bulk flow, §7.1; and the GA model, Bertschinger & Juszkiewicz 1988). However, toy modeling imposes an oversimplified geometry associated with assumed "sources" on a complex velocity field which actually arises from a continuous field of asymmetric density fluctuations (§4.5, §5). Moreover, the bulk velocity statistic computed globally suffers from a bias due to the large-scale sampling gradients. The monopole, involving the radial decrease of sampling density and rise of errors, has the effect of reducing the effective volume and thus reducing the apparent conflict between the high bulk flow and the theoretical expectations (Kaiser 1988). The sampling dipole, arising for example from oversampling in the GA direction, tends to enhance the component of the bulk velocity in that direction because it is dominated by the velocity in a smaller effective volume (Regos & Szalay 1988; Szalay 1988). The sampling quadrupole arising from the Galactic zone of avoidance (*zoa*) introduces larger shot-noise into the component parallel to the Galactic plane, which could also result in an artificially high bulk velocity. These biases can could be partially cured by equal-volume weighting (§4.2).

### 4.2. Potential Analysis

If the LSS evolved according to GI, then the large-scale velocity field is expected to be *irrotational*, $\boldsymbol{\nabla} \times \boldsymbol{v} = 0$. Any vorticity mode would have decayed during the linear regime as the universe expanded, and, based on Kelvin's circulation theorem, the flow remains vorticity-free in the quasi-linear regime as long as it is laminar. Irrotationality implies that the velocity field can be derived from a scalar potential, $\boldsymbol{v}(\boldsymbol{x}) = -\boldsymbol{\nabla}\Phi(\boldsymbol{x})$, so the radial velocity field $u(\boldsymbol{x})$ should contain enough information for a full reconstruction.

In the POTENT procedure (Bertschinger & Dekel 1989) the potential is computed by integration along radial rays from the observer

$$\Phi(\boldsymbol{x}) = -\int_0^r u(r',\theta,\phi)dr' ,\qquad(19)$$

and the two missing transverse velocity components are then recovered by differentiation. Then $\delta(\boldsymbol{x})$ is approximated by $\delta_c$ (eq. 7). The non-trivial step is the *smoothing* of the data into $u(\boldsymbol{x})$. The aim in POTENT (Dekel *et al.* 1990) is to reproduce the $u(\boldsymbol{x})$ that would have been obtained had the true $\boldsymbol{v}(\boldsymbol{x})$ been sampled densely and uniformly and smoothed with a spherical Gaussian window of radius $R_s$. With the data as available $u(\boldsymbol{x}_c)$ is taken to be the value at $\boldsymbol{x}=\boldsymbol{x}_c$ of an appropriate *local* velocity model $\boldsymbol{v}(\alpha_k,\boldsymbol{x}-\boldsymbol{x}_c)$ obtained by minimizing the sum (17) in terms of the parameters $\alpha_k$ within an appropriate local window $W_i = W(\boldsymbol{x}_i,\boldsymbol{x}_c)$ chosen as follows.

*Tensor window.* Unless $R_s \ll r$, the $u_i$s cannot be averaged as scalars because the directions $\hat{\boldsymbol{x}}_i$ differ from $\hat{\boldsymbol{x}}_c$, so $u(\boldsymbol{x}_c)$ requires a fit of a local 3D model. The original POTENT used the simplest local model, $\boldsymbol{v}(\boldsymbol{x}) = \boldsymbol{B}$, for which the solution can be expressed explicitly in terms of a tensor window function.

*Window bias.* The tensorial correction to the spherical window has conical symmetry, weighting more heavily objects of large $\hat{\boldsymbol{x}}_i \cdot \hat{\boldsymbol{x}}_c$. The resultant bias of a true infall transverse to the LOS is a flow towards the LG, *e.g.* $\sim 300$ km s$^{-1}$ at the GA in the current reconstruction. A way to reduce this bias is by generalizing $\boldsymbol{B}$ into a *linear* velocity model,

$$\boldsymbol{v}(\boldsymbol{x}) = \boldsymbol{B} + \bar{\bar{\boldsymbol{L}}} \cdot (\boldsymbol{x} - \boldsymbol{x}_c) ,\qquad(20)$$

with $\bar{\bar{\boldsymbol{L}}}$ a symmetric tensor, which ensures local irrotationality. The linear terms tend to "absorb" most of the bias, leaving $\boldsymbol{v}(\boldsymbol{x}_c) = \boldsymbol{B}$ less biased. Unfortunately, a high-order model tends to pick undesired small-scale noise. The optimal compromise is found to be a first-order model fit out to $r = 40$ h$^{-1}$Mpc, smoothly changing to a zeroth-order fit beyond 60 h$^{-1}$Mpc (Dekel *et al.* 1994).

*Sampling-gradient bias* (SG). If the true velocity field is varying within the effective window, the non-uniform sampling introduces a bias because the smoothing is galaxy-weighted whereas the aim is equal-volume weighting. One should weight each object by the local volume it "occupies", *e.g.* $V_i \propto R_n^3$ where $R_n$ is the distance to the n-th neighboring object (*e.g.* $n = 4$). This procedure is found via simulations to reduce the SG bias in Mark III to negligible levels typically out to 60 h$^{-1}$Mpc but away from the zoA. The $R_n(\boldsymbol{x})$ field can serve later as a flag for poorly sampled regions, to be excluded from any quantitative analysis.

*Reducing random errors.* The ideal weighting for reducing the effect of Gaussian noise has weights $W_i \propto \sigma_i^{-2}$ but this spoils the carefully-designed volume weighting, biasing $u$ towards its values at smaller $r_i$ and at nearby clusters where the errors are small. A successful compromise is to weight by both, *i.e.*

$$W(\boldsymbol{x}_i,\boldsymbol{x}_c) \propto V_i\,\sigma_i^{-2}\,\exp[-(\boldsymbol{x}_i-\boldsymbol{x}_c)^2/2R_s^2] .\qquad(21)$$

Note that POTENT could alternatively vary $R_s$ to keep the random errors at a constant level, but at the expense of producing fields not directly comparable to theoretical models of uniform smoothing. Another way to reduce noise is to eliminate badly-observed galaxies with large residuals $|u_i - u(\boldsymbol{x}_i)|$, where $u(\boldsymbol{x}_i)$ is obtained in a pre-POTENT smoothing stage. The whole smoothing procedure has been developed with the help of carefully-designed mock catalogs "observed" from simulations.

*Estimating the random errors.* The errors in the recovered fields are assessed by Monte-Carlo simulations, where the input distances are perturbed at random using a Gaussian of standard deviation $\sigma_i$ before being fed into POTENT. The error in $\delta$ at a grid point is estimated by the standard deviation of the recovered $\delta$ over the Monte-Carlo simulations, $\sigma_\delta$ (and similarly $\sigma_v$). In the well-sampled regions, which extend in Mark III out to 40–60 h$^{-1}$Mpc, the errors are $\sigma_\delta \approx 0.1$–$0.3$, but they may blow up in certain regions at large distances. To exclude noisy regions, any quantitative analysis should be limited to points where $\sigma_v$ and $\sigma_\delta$ are within certain limits.

Several variants of POTENT are worth mentioning. The potential integration is naturally done along radial paths using only $u(\boldsymbol{x})$ because the data are radial velocities, but recall that the smoothing procedure determines a 3D velocity using the finite effective opening angle of the window $\sim R_s/r$. The transverse components are normally determined with larger uncertainty but the non-uniform sampling may actually cause the minimum-error path to be non-radial, especially in regions where $R_s/r \sim 1$. For example, it might be better to reach the far side of a void along its populated periphery rather than through its empty center. The optimal path can be determined by a *max-flow* algorithm (Simmons *et al.* 1994). It turns out in practice that only little can be gained by allowing non-radial paths because large empty regions usually occur at large distances where the transverse components are very noisy. Still, it is possible that the derived potential can be somewhat improved by averaging over many paths.

The use of the opening angle to determine the transverse velocities can be carried one step further by fitting the data to a power-series generalizing the linear model,

$$v_i(\boldsymbol{x}) = B_i + L_{ij}\tilde{x}_j + Q_{ijk}\tilde{x}_j\tilde{x}_k + C_{ijkl}\tilde{x}_j\tilde{x}_k\tilde{x}_l + ... , \qquad (22)$$

where $\tilde{\boldsymbol{x}} = \boldsymbol{x} - \boldsymbol{x}_c$. If the matrices are all symmetric then the velocity model is automatically irrotational, and can therefore be used as the final result without appealing to the potential, and with the density being automatically approximated by $\delta_c(\boldsymbol{x}) = \|I - L_{ij}\| - 1$ (eq. 7). The fit must be local because the model tends to blow up at large distances. The expansion can be truncated at any order, limited by the tendency of the high-order terms to pick up small-scale noise. The smoothing here is not a separate preceding step, the SG bias is reduced, and there is no need for numerical integration or differentiation, but the effective smoothing is again not straightforwardly related to theoretical models of uniform smoothing (Blumenthal & Dekel, in prep.).

Another method of potential interest without a preliminary smoothing step is based on *wavelet analysis*, which enables a natural isolation of the structure on different scales (Rauzy *et al.* 1993). The effective smoothing involves no loss of information and no specific scale or shape for the wavelet, and the analysis is global and done in one step. How successful this method will be in dealing with noisy data and its comparison with theory and other data still remains to be seen.

### 4.3. Regularized Multi-Parameter Models – Wiener Filter

A natural generalization of the toy models of §4.1 is a global expansion of the fields in a discrete set of basis functions, such as a Fourier series of the sort

$$\delta = \sum_{\boldsymbol{k}}[a_{\boldsymbol{k}}\sin(\boldsymbol{k}\cdot\boldsymbol{x}) + b_{\boldsymbol{k}}\cos(\boldsymbol{k}\cdot\boldsymbol{x})], \quad \boldsymbol{v} = \sum_{\boldsymbol{k}} \frac{\boldsymbol{k}}{k^2}[a_{\boldsymbol{k}}\cos(\boldsymbol{k}\cdot\boldsymbol{x}) - b_{\boldsymbol{k}}\sin(\boldsymbol{k}\cdot\boldsymbol{x})], \quad (23)$$

with a very large number ($2m$) of free Fourier coefficients, and where $\delta$ and $\boldsymbol{v}$ are related via the linear approximation (4). The maximization of the likelihood involves a $2m \times 2m$ matrix inversion and, even if $2m$ is appropriately smaller than the number of data, the solution will most likely follow Murphy's law and blow up at large distances, yielding large spurious fluctuations where the data are sparse, noisy and weakly constraining, so the global fit requires some kind of regularization. Kaiser & Stebbins (1991) proposed to maximize the probability of the parameters subject to the data *and* an assumed *prior model* for the probability distribution of the Fourier coefficients; they assumed a Gaussian distribution with a PS $\langle a_{\boldsymbol{k}}^2 \rangle = \langle b_{\boldsymbol{k}}^2 \rangle = P_0 k^n$.

This is in fact an application of the *Wiener filter* method (*e.g.* Rybicki & Press 1992) for finding the optimal estimator of a field $\delta(\boldsymbol{x})$ which is a linear functional of another field $u(\boldsymbol{x})$, given noisy data $u_i$ of the latter and an assumed prior model. It is

$$\delta_{opt}(\boldsymbol{x}) = \langle \delta(\boldsymbol{x}) u_i \rangle \langle u_i u_j \rangle^{-1} u_j, \quad (24)$$

where the indices run over the data (Hoffman 1994; Stebbins 1994). If $u_i = u(\boldsymbol{x}_i) + \epsilon_i$ with $\epsilon_i$ independent random errors of zero mean, then the cross-correlation terms are $\langle \delta(\boldsymbol{x}) u(\boldsymbol{x}_i) \rangle$, and the auto-correlation matrix is

$$\langle u_i u_j \rangle = \langle u(\boldsymbol{x}_i) u(\boldsymbol{x}_j) \rangle + \epsilon_i^2 \delta_{ij}, \quad (25)$$

both given by the prior model. $\delta_{opt}$ is thus determined by the model where the errors dominate, and by the data where the errors are small. If the assumed prior is Gaussian, the optimal estimator is also the most probable field given the data, which is a generalization of the conditional mean given one constraint (*e.g.* Dekel 1981). This conditional mean field can be the basis for a general algorithm to produce constrained random realization (Hoffman and Ribak 1991). The same technique can also be applied to the inverse problem of recovering the velocity from observed density. Note (Lahav *et al.* 1994) that for Gaussian fields the above procedure is closely related to maximum-entropy reconstruction of noisy pictures (Gull & Daniell 1978).

The maximum-probability method has so far been applied in a preliminary way to heterogeneous data in a box of side 200 $h^{-1}$Mpc (to eliminate periodic boundary effects) with 18 $h^{-1}$Mpc resolution. No unique density field came out as different priors ($-3 \le n \le 1$) led to different fits with similar $\chi^2$. This means that the data used were not of sufficient quality to determine the fields this way and it would be interesting to see how this method does when applied to better data. The method still lacks a complete error analysis, it needs to somehow deal with non-linear effects, and it needs to correct for IM bias (perhaps as in §3.2).

A general undesired feature of maximum probability solutions is that they tend to be oversmoothed in regions of poor data, relaxing to $\delta = 0$ in the extreme case of no data. This is unfortunate because the signal of true density is modulated by the density and quality of sampling – a sampling bias which replaces the SG bias. The effectively varying smoothing length can affect any dynamical use of the reconstructed field (*e.g.* deriving a gravitational acceleration from a density field), as well as prevent a straightforward comparison with other data or with uniformly-smoothed theoretical fields. Yahil (1994) has recently proposed a modified filtering method which partly cures this problem by forcing the recovered field to have a constant variance.

### 4.4. Malmquist-Free Analysis

The selection bias (§3.2) can be practically eliminated from the calibration of an inverse TF relation, $\eta(M)$, as long as the internal velocity parameter $\eta$ does not explicitly enter the selection process. An inverse analysis requires assuming a parametric model for the velocity field, $\boldsymbol{v}(\alpha_k, \boldsymbol{x})$ (Schechter 1980). The sum minimized instead of (17) is

$$\sum_i W_i \, [\eta_i(observed) - \eta_i(model)]^2 \; , \qquad (26)$$

with the model $\eta$ given by the *inverse* TF relation,

$$\eta_i(model) = \tilde{a} + \tilde{b}M_i = \tilde{a} + \tilde{b}(m_i - 5\log r_i) \; , \qquad r_i = z_i - \hat{\boldsymbol{x}}_i \cdot \boldsymbol{v}(\alpha_k, \boldsymbol{x}_i) \; . \qquad (27)$$

The parameters are the inverse TF parameters $\tilde{a}$ and $\tilde{b}$ and the $\alpha_k$ characterizing the velocity model. This method is indeed free of inferred-distance M bias as long as $r_i$ is uniquely derived from $z_i$, and not from the inferred diatance.

An inverse method was first used by Aaronson *et al.* (1982b) to fit a Virgo-centric toy model to a local sample of spirals, and attempts to implement the inverse method to an extended sample with a general velocity model are in progress (MFPOT by Yahil *et al.* 1994). This is a non-trivial problem of non-linear multi-parameter minimization with several possible routes. If the velocity model is expressed in $z$-space, then $r_i$ is given explicitly by $z_i$ but the results suffer from oversmoothing in collapsing regions. If in $r$-space, then $r_i$ is implicit in the second equation of (27) requiring iterative minimization, *e.g.* by carrying $r_i$ from one iteration to the next one. The velocity model could be either global or local, with the former enabling a simultaneous minimization of the TF and velocity parameters and the latter requiring a sequential minimization of global TF parameters and local velocity parameters. While the results are supposed to be free of Malmquist bias, they suffer from other biases which have to be carefully diagnosed and corrected for. The inverse results will have to The inverse method is also being generalized to account for small-scale velocity noise (Willick, Burstein & Tornen 1994). The Malmquist-free results will have to be consistent with the M-corrected forward results before one can put to rest the crucial issue of M bias and its effect on our LSS results.

## 4.5. Fields of Velocity and Mass Density

Figure 3 shows supergalactic-plane maps of the velocity field in the CMB frame and the associated $\delta_c$ field (for $\Omega = 1$) as recovered by POTENT from the preliminary Mark III data. The data are reliable out to $\sim 60$ $h^{-1}$Mpc in most directions outside the Galactic plane ($Y = 0$), and out to $\sim 70$ $h^{-1}$Mpc in the direction of the GA (left-top) and Perseus-Pisces (PP, right-bottom). Both large-scale ($\sim 100$ $h^{-1}$Mpc) and small-scale ($\sim 10$ $h^{-1}$Mpc) features are important; *e.g.* the bulk velocity reflects properties of the initial fluctuations and of the DM (§7), while the small-scale variations indicate the value of $\Omega$ (§8).

The velocity map shows a clear tendency for motion from right to left, in the general direction of the LG-CMB motion ($L, B = 139°, -31°$). The bulk velocity within 60 $h^{-1}$Mpc is $300 - 350$ km s$^{-1}$ towards ($L, B \approx 166°, -20°$) (§7.1) but the flow is not coherent over the whole volume sampled, *e.g.* there are regions in front of PP and at the back of the GA where the $XY$ velocity components vanish, *i.e.* the streaming relative to the LG is opposite to the bulk flow direction. The velocity field shows local convergences and divergences which indicate strong density variations on scales about twice as large as the smoothing scale.

The GA at 12 $h^{-1}$Mpc smoothing and $\Omega = 1$ is a broad density peak of maximum height $\delta = 1.2 \pm 0.3$ located near the Galactic plane $Y = 0$ at $X \approx -40$ $h^{-1}$Mpc. The GA extends towards Virgo near $Y \approx 10$ (the "Local Supercluster"), towards Pavo-Indus-Telescopium (PIT) across the Galactic plane to the south ($Y < 0$), and towards the Shapley concentration behind the GA. The structure at the top roughly coincides with the "Great Wall" of Coma, with $\delta \approx 0.5$. The PP peak which dominates the right-bottom is centered near Perseus with $\delta = 1.0 \pm 0.4$. PP extends towards Aquarius in the southern hemisphere, and connects to Cetus near the south Galactic pole, where the "Southern Wall" is seen in redshift surveys. Underdense regions separate the GA and PP, extending from bottom-left to top-right. The deepest region in the Supergalactic plane, with $\delta = -0.7 \pm 0.2$, roughly coincides with the galaxy-void of Sculptor (Kauffman *et al.* 1991).

One can still find in the literature statements questioning the very existence of the GA (*e.g.* Rowan-Robinson 1993), which simply reflect ambiguous definitions for this phenomenon. A GA clearly exists in the sense that the dominant feature in the local inferred velocity field is a coherent convergence, centered near $X \approx -40$. It is another question whether the associated density peak has a counterpart in the galaxy distribution or is a separate, unseen entity. The GA is ambiguous only in the sense that the good correlation observed between the mass density inferred from the velocities and the galaxy density in redshift surveys is perhaps not perfect (§6.2).

Other cosmographic issues of debate are whether there exists a back-flow behind the GA in the CMB frame, and whether PP and the LG are approaching each other. These effects are detected by the current POTENT analysis only at the $1.5\sigma$ level in terms of the random uncertainty. Furthermore, the freedom in the zero-point of the distance indicators permits adding a Hubble-like peculiar velocity which can balance the GA back flow and make PP move away from the LG. Thus, these issues remain debatable.

To what extent should one believe the recovery in the ZOA which is empty of tracers? The velocities observed on the two sides of the ZOA are used as probes of the mass in the ZOA. The interpolation is based on the assumed irrotationality, where the recovered transverse components enable a reconstruction of the mass-density. However, while the SG bias can be corrected where the width of the ZOA is smaller than the smoothing length, the result could be severely biased where the unsampled region is larger. With 12 h$^{-1}$Mpc smoothing in Mark III the interpolation is suspected of being severely biased in ∼ 50% of the ZOA at $r = 40$ h$^{-1}$Mpc (where $R_4 > R_s$), but the interpolation is pretty safe in the highly populated GA region, for example. Indeed, a deep survey of optical galaxies at small Galactic latitudes recently discovered a very big cluster centered at $(l, b, z) = (325°, -7°, \sim 45$ h$^{-1}$Mpc) (R. Kraan-Korteweg, private comm.), very near the central peak of the GA as predicted by POTENT, ∼ $(320°, 0°, 40$ h$^{-1}$Mpc) (Kolatt *et al.* 1994).

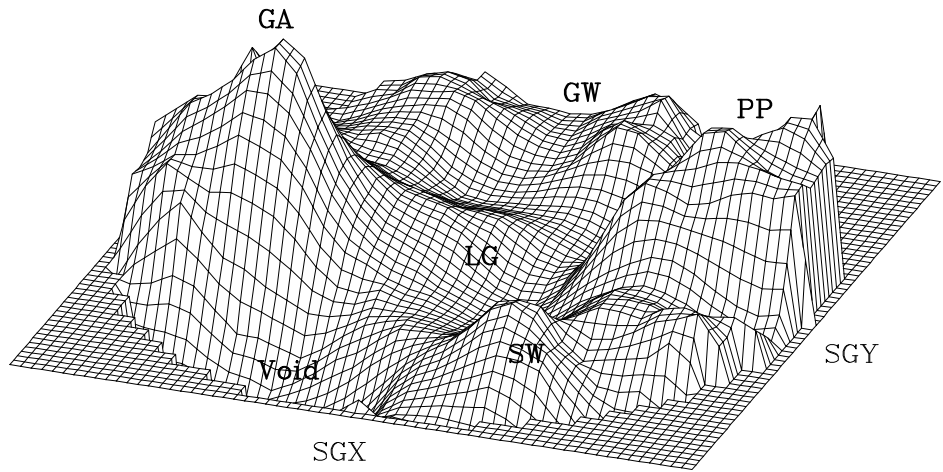

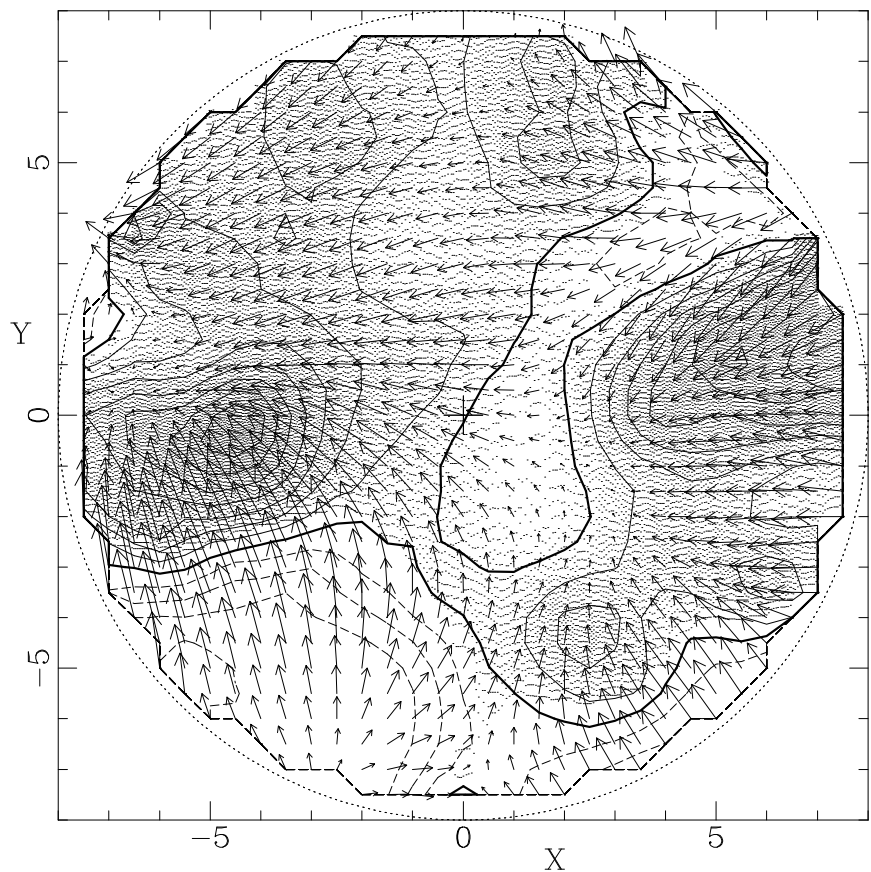

**Figure 3:** The fluctuation fields of velocity and *mass*-density in the Supergalactic plane as recovered by POTENT from the Mark III velocities of $\sim 3000$ galaxies with 12 $h^{-1}$Mpc smoothing. The vectors shown are projections of the 3D velocity field in the CMB frame. Distances and velocities are in 1000 km s$^{-1}$. Contour spacing is 0.2 in $\delta$, with the heavy contour marking $\delta = 0$ and dashed contours $\delta < 0$. The LG is at the center. The GA is on the left, PP on the right, and Coma is at the top. The grey-scale in the contour map and the height of the surface in the landscape map are proportional to $\delta$ (Dekel *et al.* 1994).

## 5. PREDICTED MOTIONS FROM THE GALAXY Z-DISTRIBUTION

All-sky complete redshift surveys provide extremely valuable data complementary to the peculiar velocity data, and efficient techniques for measuring redshifts make them deeper, denser, and more uniform. Under the assumption of GI and an assumed biasing relation between galaxies and mass, a redshift survey enables an independent reconstruction of the density and velocity fields. The comparison of the smoothed fields recovered from redshifts and from velocities is a very important tool for testing the basic hypotheses and for determining the cosmological parameters.

The solution to the linearized GI equation $\nabla \cdot \boldsymbol{v} = -f\delta$ for an irrotational field is

$$\boldsymbol{v}(\boldsymbol{x}) = \frac{f}{4\pi} \int_{\text{all space}} d^3 x' \, \delta(\boldsymbol{x}') \frac{\boldsymbol{x}' - \boldsymbol{x}}{|\boldsymbol{x}' - \boldsymbol{x}|^3} \ . \tag{28}$$

The velocity is proportional to the gravitational acceleration, which ideally requires full knowledge of the distribution of mass in space. In practice (Yahil *et al.* 1991) one is provided with a flux-limited, discrete redshift survey, obeying some radial selection function $\phi(r)$. The galaxy density is estimated by

$$1 + \delta_g(\boldsymbol{x}) = \sum n^{-1} \phi(r_i)^{-1} \delta^3_{dirac}(\boldsymbol{x} - \boldsymbol{x}_i) \ , \tag{29}$$

where $n \equiv V^{-1} \sum \phi(r_i)^{-1}$ is the mean galaxy density, and the inverse weighting by $\phi$ restores the equal-volume weighting. Eq. (28) is then replaced by

$$\boldsymbol{v}(\boldsymbol{x}) = \frac{\beta}{4\pi} \int_{r < R_{\max}} d^3 x' \, \delta_g(\boldsymbol{x}') \, S(|\boldsymbol{x}' - \boldsymbol{x}|) \frac{\boldsymbol{x}' - \boldsymbol{x}}{|\boldsymbol{x}' - \boldsymbol{x}|^3} \ . \tag{30}$$

Under the assumption of linear biasing, $\delta_g = b\delta$, the cosmological dependence enters through $\beta \equiv f(\Omega)/b$. The integration is limited to $r < R_{max}$ where the signal dominates over shot-noise. $S(\boldsymbol{y})$ is a small-scale smoothing window ($\geq 500$ km s$^{-1}$) essential for reducing the effects of non-linear gravity, shot-noise, distance uncertainty, and triple-value zones. The distances are estimated from the redshifts in the LG frame by

$$r_i = z_i - \hat{\boldsymbol{x}}_i \cdot [\boldsymbol{v}(\boldsymbol{x}_i) - \boldsymbol{v}(0)] \ . \tag{31}$$

Equations (30-31) can be solved iteratively: make a first guess for the $\boldsymbol{x}_i$, compute the $\boldsymbol{v}_i$ by eq. (30), correct the $\boldsymbol{x}_i$ by eq. (31), and so on until convergence. The convergence can be improved by increasing $\beta$ gradually during the iterations. Relevant issues follow.

*Selection function.* An accurate knowledge of the probability that a galaxy at a given distance be included in the sample is essential, especially at large distances where $\phi^{-1}$ can introduce large errors. For a given flux limit, $\phi$ can be evaluated together with the luminosity function using a maximum-likelihood technique independent of density inhomogeneities.

*Zone of Avoidance.* Regions in the sky not covered by the survey have to be filled with mock galaxies by some method of extrapolation from nearby regions. One way is to distribute these galaxies Poissonianly with the mean density of an adjacent volume, or to actually clone the adjacent region (Hudson 1993a). A more sophisticated extrapolation uses spherical harmonics and the Wiener filter method (*e.g.* Lahav *et al.* 1994).

*Triple-valued zones.* Galaxies in three different positions along a LOS through a contracting region may have the same redshift. Given a redshift in a collapsing region where the problem is not resolved by the smoothing used, one can either take some average of the three solutions, or make an intelligent choice between them, *e.g.* by using the velocity field derived from observed velocities.

*Estimating shot-noise.* This major source of error due to the finite number of galaxies can be crudely estimated using bootstrap simulations, where each galaxy is replaced with $k$ galaxies, $k$ being a Poisson deviate of $\langle k \rangle = 1$. For each realization one calculates $\phi$ and $n$, corrects for the ZOA, and solves for the linear velocity field. The mean and variance of the resulting density field are measures of the systematic and random errors. The bootstrap simulations demonstrate that the uncertainty in $\delta_g$ from the 1.2 Jy IRAS sample is typically less than 50% of the uncertainty in the density derived by POTENT from observed Mark III velocities.

*Nonlinear Biasing.* Galaxies need not be faithful tracers of the mass (*e.g.* Dekel & Rees 1987), but there is growing evidence that they are strongly correlated (§6.2). This correlation can be crudely assumed to be a deterministic relation between the local smoothed density fields, *e.g.* linear biasing $\delta_g = b\delta$, which is one realization of the linear statistical relation between the variances of the fields predicted for linear density peaks in a Gaussian field (Kaiser 1984; Bardeen *et al.* 1986). However, a more sophisticated analysis may require a more realistic biasing relation, *e.g.* an exception from linear biasing which must be made for negative $\delta_g$ and $b < 1$ to prevent $\delta$ from falling unphysically below $-1$. The non-linear generalization $1 + \delta_g = (1 + \delta)^b$ is useful (*e.g.* Dekel *et al.* 1993) and it fits quite well the biasing seen in simulations of galaxy formation in a CDM scenario involving cooling and gas dynamics (Cen & Ostriker 1993), except that a small correction is needed to force the means of $\delta$ and $\delta_g$ to vanish simultaneously as required by definition.

*Quasilinear correction.* Even after $12 \, h^{-1} \text{Mpc}$ smoothing, $\delta_g$ is of order unity in places, necessitating a quasi-linear treatment. Local approximations from $\boldsymbol{v}$ to $\delta$ were discussed in §2, but the non-local nature of the inverse problem makes it less straightforward. A possible solution is to find an inverse relation of the sort $\boldsymbol{\nabla v} = F(\Omega, \delta_g)$, including non-linear biasing and non-linear gravity. This is a Poisson-like equation in which $-\beta \delta_g(\boldsymbol{x})$ is replaced by $F(\boldsymbol{x})$, and since the smoothed velocity field is irrotational for quasi-linear perturbations as well, it can be integrated analogously to eq. (30). With smoothing of $10 \, h^{-1} \text{Mpc}$ and $\beta = 1$, the approximation based on $\delta_c$ has an *rms* error $< 50 \, \text{km s}^{-1}$ (Mancinelli *et al.* 1994). Note that for very small $b$ the $\delta$ associated with the observed $\delta_g$ could be non-linear to the extent that the quasi-linear approximations break down. When $\delta$ is given, the Poisson equation can be integrated more efficiently by using grid-based FFT techniques than by straightforward summation. The $r$-space $\delta(\boldsymbol{x})$ deduced from the $z$-space galaxy distribution is not too sensitive to non-linear effects or the value of $\Omega$, so it is a reasonable shortcut to correct for non-linear effects only in the transformation from $\delta$ to $\boldsymbol{v}$ using FFT.

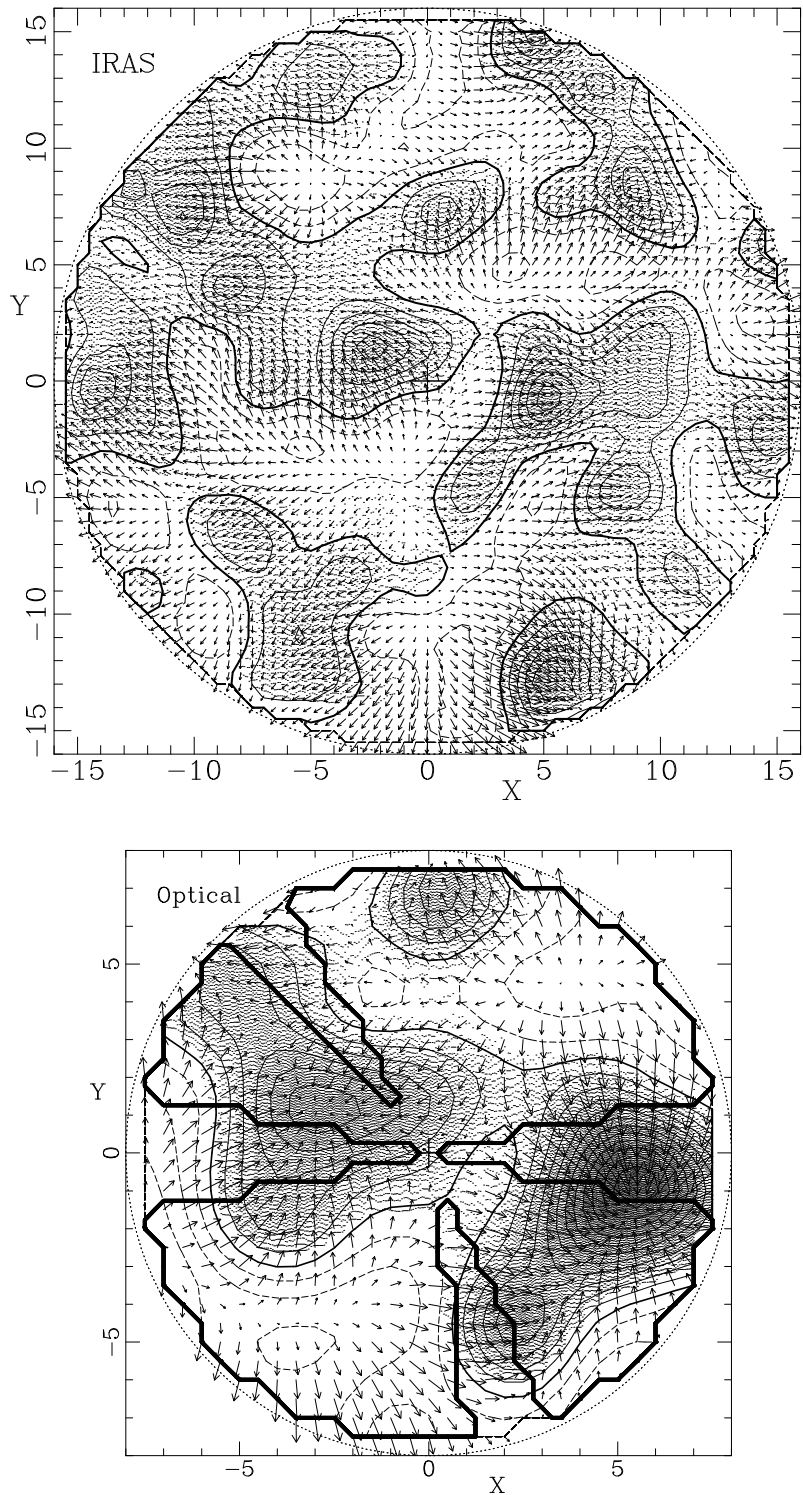

**Figure 4:** The fluctuation fields of *galaxies* in the Supergalactic plane as deduced from redshift surveys with 10 $h^{-1}$Mpc smoothing. Distances and velocities are in km s$^{-1}$. Density contour spacing is 0.2. (a) Reconstructed by A. Yahil and M. Strauss using a power-preserving filter from the IRAS 1.2 Jy data in a sphere of radius 160 $h^{-1}$Mpc. (b) Reconstructed by Hudson (1993a, 1994) from optical data within 80 $h^{-1}$Mpc, extrapolated into the unsampled areas ouside the heavy contour.

The IRAS Point Source Catalog served as the source for two very valuable redshift surveys, which have been carried out and analyzed in parallel. One contains the 5313 galaxies brighter than 1.2 Jy at $60\mu$ with sky coverage (almost) complete for $|b| > 5°$ covering 88% of the sky (Strauss *et al.* 1990a, 1992, to 1.9 Jy; extended by Fisher 1992). The other is a 1-in-6 sparsely-sampled survey of $\approx 2300$ galaxies down to the IRAS flux limit of 0.6 Jy (Rowan-Robinson *et al.* 1990, QDOT), which is now being extended to a fully-sampled survey (Saunders *et al.* in prep.). As for optical galaxies, Hudson (1993a) developed a clever way to reconstruct a statistically uniform density field out to $\sim 80\ h^{-1}$Mpc by combining the UGC/ESO diameter-limited angular catalogs and the ZCAT incomplete redshift survey. Figure 4 shows maps of the galaxy density fields and the associated predicted velocity fields, with the main features corresponding to those recovered from observed velocities (Fig. 3), *e.g.* the GA, PP and Coma superclusters and the voids in between (§6.2).

## 6. TESTING BASIC HYPOTHESES

The data and analyses described above, combined with other astrophysical data, can help us evaluate some of the basic hypotheses laid out in the Introduction.

### 6.1. CMB Fluctuations versus Motions

If the CMB defines a standard cosmological frame then the established helio-centric dipole pattern of $\delta T/T = (1.23 \pm 0.01) \times 10^{-3}$ is a direct measurement of peculiar velocity of the LG: $V(0) = 627 \pm 22$ km s$^{-1}$ towards $(l, b) = (276° \pm 3°, +30° \pm 3°)$ (Kogut *et al.* 1993). The Copernican hypothesis then implies that large peculiar velocities *exist* in general, and the question is only how *coherent* are they. A more esoteric interpretation tried to explain the CMB dipole by a horizon-scale gradient in entropy, relic of bubbly inflation, which can be made consistent with the smallness of the quadrupole and the achromaticity of the dipole (*e.g.* Gunn 1988; Paczynski & Piran 1990). In addition to the general objection to global anisotropy based on the simplicity principle of Occam's razor, a non-velocity interpretation fails to explain the fact that the gravitational acceleration vector at the LG, $\boldsymbol{g}(0)$, as inferred from the galaxy distribution in our cosmological neighborhood (§8.1), is within 20° of $\boldsymbol{V}(0)$ and of a similar amplitude of several hundred km s$^{-1}$ for any reasonable choice of $\Omega$. This argument in favor of the reality of $\boldsymbol{V}(0)$ and the standard GI picture is strengthened by the fact that a similar $\boldsymbol{g}(0)$ is obtained from the POTENT mass field derived from velocity divergences in the neighborhood (Kolatt *et al.* 1994).

The measurements of CMB fluctuations on scales $\leq 90°$ are independent of the local streaming motions, but GI predicts an intimate relation between them. The CMB fluctuations are associated with fluctuations in gravitational potential, velocity and density in the surface of last scattering at $z \sim 10^3$, while similar fluctuations in our neighborhood have grown by gravity to produce the dynamical structure observed. The comparison between the two is therefore a crucial test for GI. Before COBE, the streaming velocities were used to predict the expected level of CMB fluctuations. The local surveyed region of $\sim 100\ h^{-1}$Mpc corresponds to a $\sim 1°$ patch on the last-scattering surface. The major effect on scales $> 1°$ is the Sachs-Wolfe effect (1967), where potential fluctuations $\Delta\Phi_g$ induce temperature fluctuations via gravitational redshift,

$$\frac{\delta T}{T} = \frac{\Delta\Phi_g}{3c^2} \sim \frac{Vx}{3c^2} \ . \tag{32}$$

Since the velocity potential is proportional to $\Phi_g$ in the linear and quasi-linear regimes, $\Delta\Phi_g$ is $\sim Vx$, where $x$ is the scale over which the bulk velocity is $V$. A typical bulk velocity of $\sim 300$ km s$^{-1}$ across $\sim 100\ h^{-1}$Mpc corresponds to $\delta T/T \sim 10^{-5}$ at $\sim 1°$. If the fluctuations are scale-invariant ($n = 1$), then $\delta T/T \sim 10^{-5}$ is expected on all scales $> 1°$. Bertschinger *et al.* (1990) produced a $\delta T/T$ map of the local region as seen by a distant observer, and predicted $\delta T/T \sim 10^{-5}$ from the local potential well associated with the GA. Now that CMB fluctuations of $\approx 10^{-5}$ have been detected in the range $1° - 90°$ (*e.g.* Smoot *et al.* 1992; Schuster *et al.* 1993; Cheng *et al.* 1993), the argument can be reversed: if one assumes GI then the *expected* bulk velocity in the surveyed volume is $\sim 300$ km s$^{-1}$, *i.e.* the motions are real. If, alternatively, one accepts the velocities as real, then the CMB-POTENT agreement is relatively a sensitive test of GI, truly addressing the specific *time-evolution* of structure predicted by GI.

### 6.2. Galaxies versus Dynamical Mass

The theory of GI plus the assumption of linear galaxy biasing predicts a correlation between the dynamical density field and the galaxy density field, which can be addressed quantitatively based on the estimated errors in the two data sets. Figure 5 compares density maps in the Supergalactic plane for IRAS 1.2 Jy galaxies and POTENT Mark III mass, both Gaussian smoothed with radius 12 $h^{-1}$Mpc. The correlation is evident – the GA, PP, Coma and the voids all exist both as dynamical entities and as structures of galaxies. A quantitative comparison of these new data is in progress, but so far an elaborate statistical analysis (Dekel *et al.* 1993) has been applied only to the earlier POTENT reconstruction based on Mark II data and IRAS 1.9 Jy survey. Noise considerations in POTENT limited that analysis to a volume $\sim (53\ h^{-1}\text{Mpc})^3$ containing $\sim 12$ independent density samples. Monte-Carlo noise simulations showed that the POTENT density is consistent with being a noisy version of the IRAS density, *i.e.* the data are in agreement with the hypotheses of GI plus linear biasing.

What is it exactly that one can learn from the observed correlation (Babul *et al.* 1994)? First, it is hard to invoke any reasonable way to make the galaxy distribution and the TF measurements agree so well unless the velocities are real (H6), provided that IM bias has been properly corrected (§3.2, §4.4). Then, it is true that gravity is the only long-range force which could attract galaxies to stream toward density concentrations, but the fact that this correlation is predicted by GI plus linear biasing does not necessarily mean that it can serve as a sensitive test for either. Recall that converging (or diverging) flows tend to generate density hills (or valleys) simply as a result of mass conservation, independent of the source of the motions.

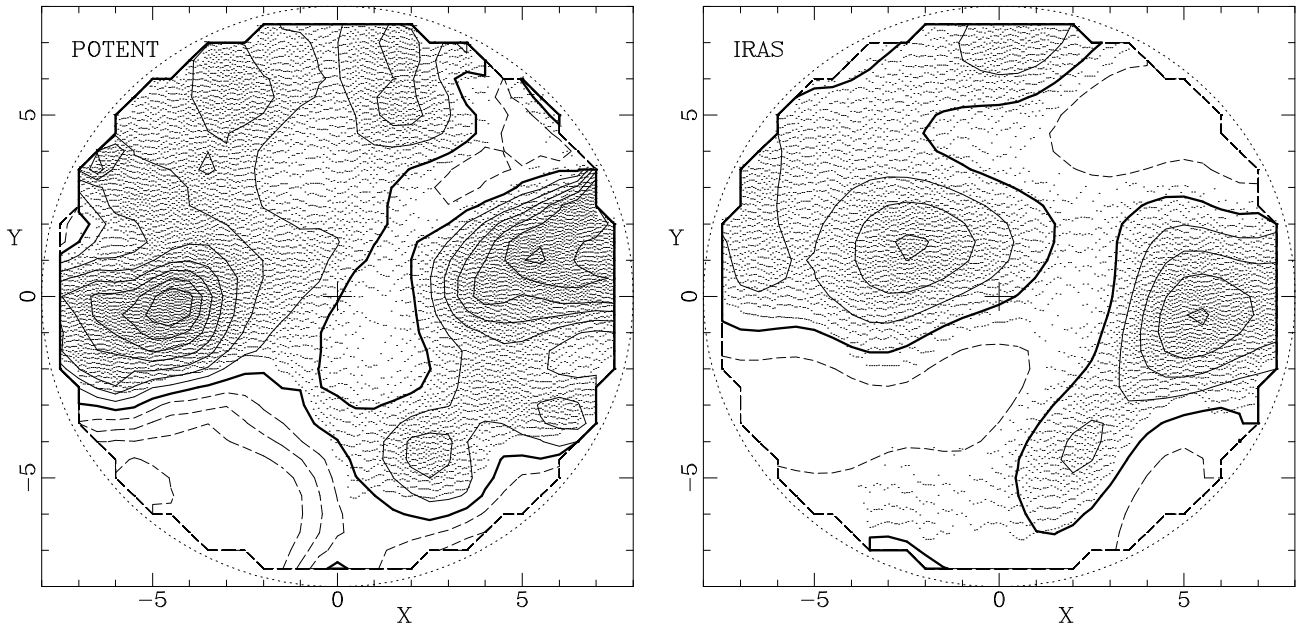

**Figure 5:** POTENT mass versus IRAS galaxy density fields in the Supergalactic plane, Gaussian smoothed with radius 12 h$^{-1}$Mpc. Contour spacing is 0.2 (Dekel *et al.* 1993).

Let us assume for a moment that galaxies trace mass, *i.e.* the linearized continuity equation, $\dot{\delta} = -\boldsymbol{\nabla}\cdot\boldsymbol{v}$, is valid for the galaxies as well. The observed correlation means $\delta \propto -\boldsymbol{\nabla}\cdot\boldsymbol{v}$, and together they imply that $\dot{\delta} \propto \delta$, or equivalently that $\boldsymbol{\nabla}\cdot\boldsymbol{v}$ is proportional to its time average. This property is not exclusive to GI; one can construct a counterexample where the velocities are produced by a non-GI impulse. Even irrotationality does not follow from $\delta \propto -\boldsymbol{\nabla}\cdot\boldsymbol{v}$; it has to be adopted based on theoretical arguments in order to enable reconstruction from radial velocities or from observed densities. Once continuity and irrotationality are assumed, the observed $\delta \propto -\boldsymbol{\nabla}\cdot\boldsymbol{v}$ implies a system of equations which is identical in all its *spatial* properties to the equations of GI, but can differ in the constants of proportionality and their temporal behavior. It is therefore impossible to distinguish between GI and a non-GI model which obeys continuity plus irrotationality based only on snapshots of *present-day* linear fluctuation fields. This makes the relation between CMB fluctuations and velocities an especially important test for GI. On the other hand, the fact that the constant of proportionality in $\delta \propto -\boldsymbol{\nabla}\cdot\boldsymbol{v}$ is indeed the same everywhere is a non-trivial requirement from a non-GI model. A version of the explosion scenario (Ostriker & Cowie 1981; Ikeuchi 1981), which tested successfully both for irrotationality and $\boldsymbol{v}-\delta$ correlation, requires certain synchronization among the explosions (Babul *et al.* 1994).

While the sensitivity of the $\boldsymbol{v}-\delta$ relation to GI is only partial, this relation turns out to be quite sensitive to the validity of a *continuity*-like relation for the *galaxies*; when the latter is strongly violated all bets are off for the $\boldsymbol{v}-\delta_g$ relation. A non-linear biasing scheme would make continuity invalid for the galaxies, which would ruin the $\boldsymbol{v}-\delta_g$ relation even if GI is valid. The observed correlation is thus a sensitive test for density *biasing*. It implies, subject to the errors, that the $\sim 10$ h$^{-1}$Mpc-smoothed densities of galaxies and mass are related via a simple, nearly linear biasing relation with $b$ not far from unity.

## 6.3. Environmental Effects: Ellipticals versus Spirals

A priori, it is possible that the inferred motions are just a reflection of systematic variations in the distance indicators, *e.g.* due to environmental variations in intrinsic galaxy properties, or in the apparent quantities due to Galactic absorption (*e.g.* Silk 1989; Djorgovski *et al.* 1989). Efforts to detect correlations between velocities and certain galaxy or environmental properties have led so far to null or at most minor detections (Aaronson & Mould 1983; Lynden-Bell *et al.* 1988; Burstein 1990a, 1990b; Burstein *et al.* 1990) – no correlation with absorption and with local galaxy density, and marginal correlations with absolute luminosity and with stellar population (Gregg 1993). Admittedly, these null results are only indicative as one cannot rule out a correlation with some other property not yet tested for.

Qualitative comparisons of the velocities of ellipticals (Es) and spirals (Ss) indicated general agreement (Burstein *et al.* 1990), and now the Mark III data enable a quantitative comparison at the same positions (Kolatt & Dekel 1994a). Figure 6 compares the two fields as derived independently by POTENT, showing a general resemblance. The radial velocities of each type were interpolated into a smoothed field $u(\boldsymbol{x})$ on a grid using 12 $h^{-1}$Mpc POTENT smoothing (§4.2), and then compared within a volume limited by the poor sampling of Es to $\simeq (50~h^{-1}\text{Mpc})^3$, containing $\sim 10$ independent sub-volumes. The two fields were found to be *consistent* with being noisy versions of the same underlying field, while the opposite hypothesis of complete independence is strongly ruled out, indicating that the consistency is not dominated by errors. A possible discrepancy indicated earlier by the Mark II data (Bertschinger 1991) is now gone with the improved data and bias-correction. The strength of this test will improve further as the E sample grows in extent (*e.g.* project EFAR by Colless *et al.* 1993) and in accuracy (*e.g.* the SBF method).

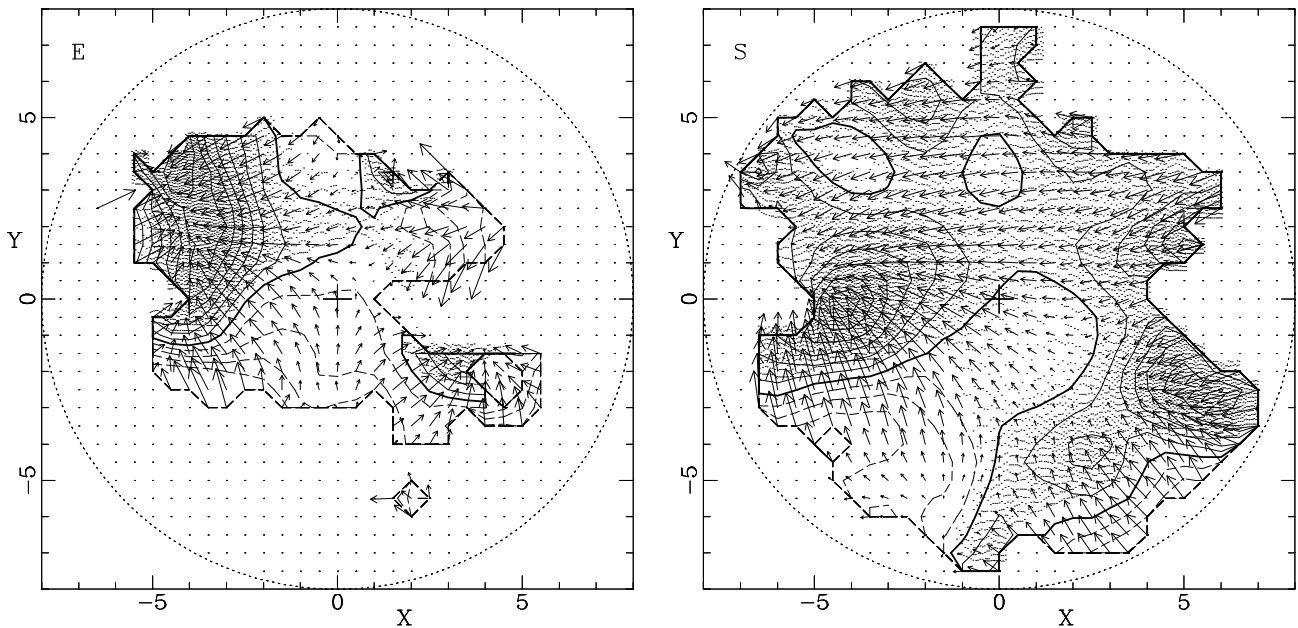

**Figure 6:** Ellipticals versus Spirals. Mass-density and velocity fields in the Supergalactic plane as recovered by POTENT independently for the two types. Smoothing is 12 $h^{-1}$Mpc. Contour spacing is 0.2. Distances and velocities are in 1000 km s$^{-1}$ (Kolatt & Dekel 1994).

The E-S correlation is consistent with motions (H6), but it cannot rule out environmental effects beyond making this idea less plausible. If such effects were dominant, then the two distance indicators would have to vary coherently, which would require that the large-scale properties of the environment be different from the local properties determining the galaxy type (E in clusters, S in field). It would also require that the properties affecting the virial equilibrium which determines the $D_n$–$\sigma$ relation, and those affecting the additional constraint involved in the TF relation at galaxy formation (§2.1), varied together in space. The one quantity that could plausibly affect the two in a correlated way is $M/L$, but to test for this one will have to compare the TF/$D_n$–$\sigma$ velocities to those inferred from an independent indicator, e.g. SBF.

The E-S correlation is consistent with the hypothesis of GI (H4) because, following Galileo, the velocities of all test bodies in a given gravitational potential are predicted to be the same as long as they share the same initial conditions. However, the observed E-S correlation does not rule out any non-GI model where all objects obtain the same velocities independent of their type, e.g. cosmological explosions or radiation pressure instabilities (Hogan & White 1986).

A practical use of the independent derivation of $u(\boldsymbol{x})$ for Ss and Es is in matching the zero points of the distance indicators, otherwise determined in an arbitrary way (§2). A 5% Hubble-like outflow has to be added to the peculiar velocities of Es from Mark II to optimally match the S data of Mark III.

## 7. THE INITIAL FLUCTUATIONS

Having assumed GI, the LSS can be traced backward in time to recover the initial fluctuations and constrain statistics which characterize them as a random field, e.g. the power spectrum (PS), and the probability distribution functions (PDFs). "Initial" here may refer either to the *linear* regime at $z \sim 10^3$ after the onset of the self-gravitating matter era, or to the origin of fluctuations in the early universe before being filtered on sub-horizon scales during the plasma-radiation era. The PS is filtered on scales $\leq 100$ h$^{-1}$Mpc by DM-dominated processes, but its shape on scales $\geq 10$ h$^{-1}$Mpc is not affected much by recent non-linear effects (because the rapid density evolution in superclusters roughly balances the slow evolution in voids at the same wavelength). The shape of the one-point PDF, on the other hand, is expected to survive the plasma era unchanged but it develops strong skewness even in the mildly-non-linear regime. Thus, the present day PS can be used as is to constrain the origin of fluctuations (large scale) and the nature of the DM (small scale), while the PDF needs to be traced back to the linear regime first. (Review: Efstathiou 1990.)

### 7.1 Power Spectrum – Dark Matter

The competing LSS formation scenarios are reviewed in Peebles (1993, §25). If the DM is all baryonic, then by nucleosynthesis constraints (see Kolb & Turner 1990, §4) the universe must be of low density, $\Omega < 0.2$, and a viable model for LSS is the Primordial Isocurvature Baryonic model (PIB) with several free parameters, typically of large relative power on large scales. With $\Omega \sim 1$ the non-baryonic DM constituents are either "hot" or "cold", and the main competing models are CDM, HDM, and MDM – a 3:7 mixture of the two (*e.g.* Blumenthal *et al.* 1988; Davis *et al.* 1992; Klypin *et al.* 1993). The main difference in the DM effect on the PS arises from free-streaming damping of the "hot" component of fluctuations on galactic scales.

*Bulk velocity.* A simple and robust statistic related to the PS is the amplitude $V$ of the vector average of the smoothed velocity field, $\boldsymbol{v}$, over a volume defined by a normalized window function $W_R(\boldsymbol{r})$ of a characteristic scale $R$ (*e.g.* top-hat),

$$\boldsymbol{V} \equiv \int d^3x\, W_R(\boldsymbol{x})\, \boldsymbol{v}(\boldsymbol{x})\ , \quad \langle V^2 \rangle = \frac{f^2}{2\pi^2} \int_0^\infty dk\, P(k)\, \tilde{W}_R^2(\boldsymbol{k})\ . \qquad (33)$$

$\langle V^2 \rangle$ is predicted for a linear model with a density spectrum $P(k)$, where $\tilde{W}_R^2(\boldsymbol{k})$, the Fourier transform of $W_R(\boldsymbol{r})$, emphasizes waves $\geq R$. The bulk velocity is obtained from the observed radial velocities by minimizing eq. (17). The report by Dressler *et al.* (1987) for the 7S Es sampled within $\sim 60$ h$^{-1}$Mpc was $V = 599 \pm 104$ towards $(l,b) = (312°, +6°)$, which was interpreted prematurely as being in severe excess of the predictions of common theories. However, this measurement cannot be directly compared to the predictions for a top-hat sphere because the effective window is much smaller due to the nonuniform sampling (SG) and weighting (Kaiser 1988). The SG bias can be crudely corrected by volume weighting as in POTENT (§4.2), at the expense of large noise. Courteau *et al.* (1993) find for the tentative Mark III data: $V_{40} = 466 \pm 48$ $(312°, +15°)$ and $V_{60} = 398 \pm 47$ $(305°, +14°)$, where $V_R$ refers to a top-hat sphere of radius $R$ h$^{-1}$Mpc. Alternatively, $\boldsymbol{V}$ can be computed from the POTENT $\boldsymbol{v}$ field by simple vector averaging from the grid. Figure 7 shows two results, one minimizing the SG bias by $V_i$ weighting, and the other reducing the random errors by weighting $\propto \sigma_i^{-2}$ (§4.2). $V_{60}$ is found to be in the range $270 - 360$ km s$^{-1}$ $(296°, +11°)$ – smaller than previous estimates. The additional random error from Monte-Carlo noise simulations is typically 15%, not including cosmic scatter due to the fact that only one sphere has been sampled.

*Mach number.* The bulk velocity is robust but it relates to the normalization of the PS, not predicted from first principles by any of the competing theories but rather normalized by some other uncertain observation, *e.g.* the CMB fluctuations or the galaxy distribution with an unknown biasing factor. A statistic which measures the *shape* of the PS free of its normalization is the cosmic Mach number (Ostriker & Suto 1989), defined as $\mathcal{M} \equiv V/S$, where $S$ is the *rms* deviation of the local velocity from the bulk velocity,

$$S^2 \equiv \int d^3x\, W_R(\boldsymbol{x})\, [\boldsymbol{v}(\boldsymbol{x}) - \boldsymbol{V}]^2\ , \quad \langle S^2 \rangle = \frac{f^2}{2\pi^2} \int_0^\infty dk\, P(k)\, [1 - \tilde{W}_R^2(\boldsymbol{k})]\ . \qquad (34)$$

$\mathcal{M}(R, R_s)$ measures the ratio of power on large scales $\gtrsim R$ to power on small scales $\gtrsim R_s$. Strauss *et al.* (1993) derived $\mathcal{M} = 1.0$ for the local Ss (Aaronson *et al.* 1982a) with $R \sim 20$ h$^{-1}$Mpc and $R_s \to 0$, and found $\sim 5\%$ of their CDM simulations to have $\mathcal{M}$ as large – a marginal rejection or consistency depending on taste. On larger scales, using POTENT with $R = 60$ h$^{-1}$Mpc (top-hat) and $R_s = 12$ h$^{-1}$Mpc (Gaussian), the tentative Mark III data yield $\mathcal{M} \sim 1$, which roughly coincides with the *rms* expected from CDM but has only $\sim 5\%$ probability to be that low for a PIB spectrum ($\Omega = 0.1$, $h = 1$, fully ionized, normalized to $\sigma_8 = 1$) (Kolatt *et al.*, in prep.).

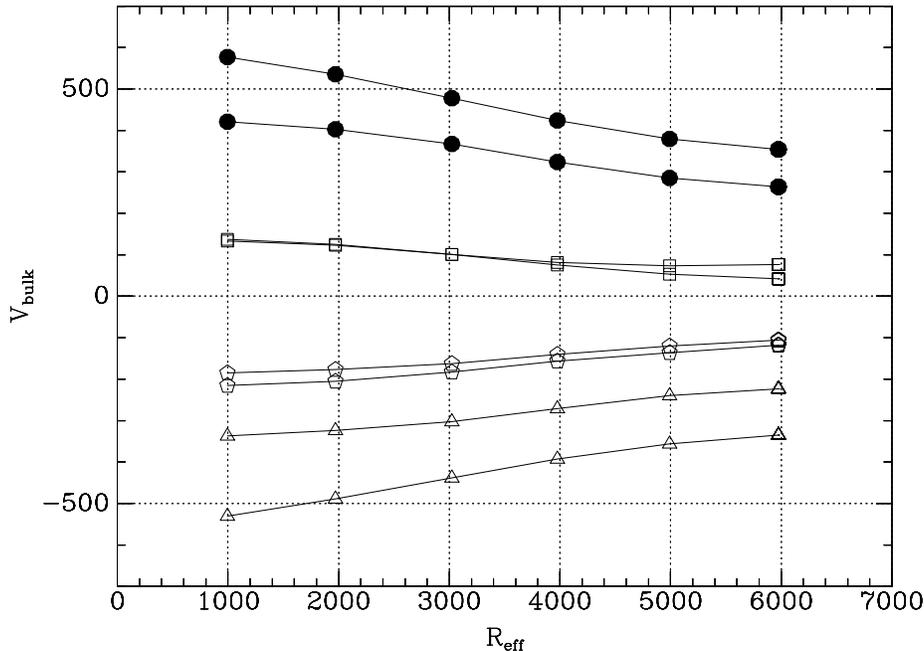

**Figure 7:** The bulk velocity in a top-hat sphere of radius $R_{eff}$ about the LG as recovered by POTENT from the Mark III data. Shown are $|V|$ (filled), $V_x$ (triangles), $V_y$ (squares), and $V_z$ (hexagons). Distances and velocities are in $\mathrm{km\,s^{-1}}$. The two results shown reflect the systematic uncertainty. The $1\sigma$ uncertainty due to random distance errors is $\simeq 15\%$ (Dekel et al. 1994).

*Power spectrum.* The velocity field by POTENT enables a preliminary determination of the mass PS itself in the range $10-100\ \mathrm{h^{-1}Mpc}$ (Kolatt & Dekel 1994b). It is best determined by the potential field, which is smoother and less sensitive to non-linear effects than its derivatives $\boldsymbol{v}$ and $\delta$. The PS was compared with the predictions of theoretical models which were N-body simulated, "observed", and fed into POTENT before the PS was computed using the same procedure applied to the observed data, thus eliminating systematic errors and also estimating random errors. The preliminary results indicate that the shape of the PS in the limited range sampled resembles a CDM-like PS with a shape parameter $\Gamma \sim 0.5$ (referring in CDM to $\Omega h$), with the power index bending toward $n \simeq 0$ by $100\ \mathrm{h^{-1}Mpc}$. The normalization for the mass PS, also obtained by Seljak & Bertschinger (1994), is $f\sigma_8 = 1.3 \pm 0.3$. This is not too sensitive to the PS shape, but still note that it mostly determined by data at $\sim 30\ \mathrm{h^{-1}Mpc}$. With $\sigma_{8o} \approx 1$ for optical galaxies, this implies $\beta_o \approx 1.3 \pm 0.3$ (compare §8). If $\Omega = 1$, the quadrupole in $\delta T/T$ by COBE corresponds to $\sigma_8 = 1.0, 0.6, 0.5$ with $\sim 45\%$ error (including cosmic scatter) for CDM, MDM and HDM spectra (yet to be calculated for an open universe), i.e. CDM is fine while HDM and MDM are $\sim 2\sigma$ low. In comparison, the PS of the different luminous objects are all well described by a CDM-like PS with $\Gamma \approx 0.25$, with the relative bias factors for Abell clusters, optical galaxies and IRAS galaxies in the ratios $4.5 : 1.3 : 1$ ($\pm 6\%$ *rms*, e.g. Peacock & Dodds 1994). If indeed $\sigma_{8o} \sim 1$ for optical galaxies, then IRAS galaxies are slightly anti-biased and $\beta_I \approx 1.3\beta_o$.

*A discrepancy on a very large scale?* The galaxy velocity field in the local $\sim 60\ \mathrm{h^{-1}Mpc}$ which has been studied quite accurately seems to be in general agreement with other observations and with the predictions of the common GI scenarios, but there is a disturbing hint for a possible discrepancy with our basic hypotheses on larger scales. Lauer and Postman (1993, LP) measured the bulk flow of the volume-limited system of 119 Abell/ACO clusters within $z < 150$ $h^{-1}$Mpc, estimating distances from luminosity versus slope of the surface-brightness profile in brightest cluster galaxies (BCG) with claimed 16% error. The LG motion with respect to this system is found to be $561 \pm 284$ km s$^{-1}$ towards $(l, b) = (220°, -28°)$, roughly 80° off the direction of the LG-CMB velocity $(276°, +30°)$. The inferred bulk velocity of the cluster system (of effective radius $\sim 80 - 110$ $h^{-1}$Mpc) relative to the CMB is $V = 689 \pm 178$ km s$^{-1}$ towards $(343°, +52°)$ ($\pm 23°$), in the general direction of the LG motion, the galaxy bulk flow, the GA and the background Shapley concentration. Taken at face value this is a *big* velocity over a *big* scale; the *rms* bulk velocity within $R = 100$ $h^{-1}$Mpc as predicted by the common theories normalized to COBE is below 200 km s$^{-1}$. However, the errors are big too!

The data have been subjected by LP to careful statistical tests. With a few tens of points contributing to the velocity in each direction, the shot-noise can clearly contribute a false signal of several hundred km s$^{-1}$. For assessing the theoretical implications, the actual observing scheme of LP was applied to N-body simulations of several competing scenarios (Strauss *et al.* 1994). Clusters were placed at the 119 highest density peaks of appropriate mass, and Galactic extinction and observational errors were modeled. As noted by LP, the fact that the measured velocity vector lies away from the *ZOA* increases the statistical significance of this measurement. Taking the error ellipsoid into account, the probability by means of $\chi^2$ of the LP result in the model universes simulated is 2.6-5.8%. Feldman & Watkins (1994), using a different technique, find the probabilities to be 6-10%. The LP bulk velocity is thus a $\sim 2\sigma$ deviation from several common theories. This is probably not enough for a serious falsification of these models, but it is certainly an intriguing result which motivates more accurate investigations. The BCG method is limited to one measurement per cluster, so an obvious strategy to reduce the errors would be to collect many TF and $D_n$–$\sigma$ distances per cluster.

A true $\sim 700$ km s$^{-1}$ velocity at $R \sim 100$ $h^{-1}$Mpc would be in serious conflict with GI. First, if this velocity is typical then it predicts larger CMB fluctuations than observed on $\sim 2°$. Second, the gravitational acceleration on the LP sphere as estimated from the spatial distribution of clusters on even larger scales (§8.1, Scaramella *et al.* 1991) predicts a flow of only $\sim 200$ km s$^{-1}$. The way to interpret the LP result in the context of the conventional theories is either as a $\sim 2$–$2.5\sigma$ statistical fluke, or as a biased result due to a yet-unresolved systematic error in the BCG method or in the sample, *e.g.* a significant velocity biasing of the clusters (which is contrary to the naive expectations from GI, however).

## 7.2 Back in Time

The forward integration of the GI equations by analytic approximations or by N-body simulations cannot be simply reversed despite the time reversiblity of gravity. It is especially hopeless in collapsed systems where memory has been erased, but the case is problematic even for linear systems. When attempting backwards integration, the decaying modes (*e.g.* Peebles 1993, §5), having left no detectable trace at present, $t_0$, would amplify noise into dominant spurious fluctuations at early times. This procedure has a negligible probability of recovering the very special initial state of almost uniform density and tiny velocities which we assume for the real universe at $t_i$. The problem is of mixed boundary conditions: some of the six phase-space variables per particle are given at $t_i$ and some at $t_0$. This problem can be solved either by eliminating the decaying modes or by applying the principle of least action.

*Zel'dovich time machines.* If the velocity field is irrotational, the Euler equation (2) can be replaced by the Bernoulli equation for the potentials, $\dot{\Phi}_v - (\nabla \Phi_v)^2/2 = -2H\Phi_v + \Phi_g$. The Z approximation, restricted to the growing mode, requires that each side vanish: one side relates $\Phi_v$ and $\Phi_g$ linearly and the other is the "Zel'dovich-Bernoulli" (ZB) equation (Nusser & Dekel 1992),
$$\dot{\varphi}_v - (\dot{D}/2)(\nabla \varphi_v)^2 = 0 \;, \tag{35}$$
with the potentials in units of $a^2 \dot{D}$. The ZB equation can be easily integrated backwards with a guaranteed uniform solution at $t_i$. $\varphi_v$ at $t_0$ is extractable from observations of velocities (§4) or galaxy density (§5), and the initial $\boldsymbol{v}$ and $\delta$ can be derived from the initial $\varphi_v$ using linear theory. While the ZB approximation conserves momentum (like $\delta_c$) one can alternatively satisfy continuity under the Z approximation (like $\delta_0$), and obtain a second-order equation for $\varphi_g$ which can be more accurate than ZB for certain purposes (Gramman 1993b, eq. 2.24, 2.25). The recovered initial $\delta_i$ has deeper valleys and shallower hills compared to naive recovery using linear theory, *e.g.* the GA is less eccentric than assumed by Juszkiewicz & Bertschinger (1988).

*Recovering the IPDF.* An important issue is whether the initial fluctuations were Gaussian or not. A Gaussian field is characterized by the joint PDFs of all order being generalized Gaussians (cf., Bardeen *et al.* 1986), and in particular the one-point probability of $\delta$ is $P(\delta) \propto \exp[-\delta^2/(2\sigma^2)]$. Common Inflation predicts Gaussian fluctuations but non-Gaussian fluctuations are allowed by certain versions of Inflation (cf., Kofman *et al.* 1990) and by models where the perturbations are seeded by cosmic strings, textures, or explosions (see Peebles 1993, §16). The present density PDF develops a log-normal shape due to non-linear effects (Coles & Jones 1991; Kofman *et al.* 1994): the tails become positively skewed because peaks collapse to large densities while the density in voids cannot become negative, and the middle develops negative skewness as density hills contract and valleys expand. On the other hand, the PDF of present-day velocity components is *in*sensitive to quasi-linear effects (Kofman *et al.* 1994).

The observed PDFs today agree with $N$-body simulations of Gaussian initial conditions (Bouchet *et al.* 1993), but they have only limited discriminatory power against initial non-Gaussianities; the development of a density PDF with a general log-normal shape may occur even in certain cases of non-Gaussian initial fluctuations (*e.g.* Weinberg & Cole 1992), and the velocity PDF becomes Gaussian under general conditions due to the central limit theorem whenever the velocity is generated by several independent density structures. A more effective strategy seems to be to take advantage of the *full* dynamical fields at $t_0$, trace them back in time, and use the linear fields to discriminate between theories. The Eulerian Z approximation can be used to directly recover the initial PDF (IPDF) as follows (Nusser & Dekel 1993). The tensor $\partial v_i/\partial x_j$ derived from $\boldsymbol{v}(\boldsymbol{x})$ is transformed to Lagrangian variables $\boldsymbol{q}(\boldsymbol{x})$. The corresponding eigenvalues $\mu_i \equiv \partial v_i/\partial x_i$ and $\lambda_i \equiv \partial v_i/\partial q_i$ are related via the key relation
$$\lambda_i = \mu_i/(1 - f^{-1}\mu_i) \;. \tag{36}$$

In the Z approximation $\bm{v} \propto \dot{D}$ so the Lagrangian derivatives $\lambda_i$ are traced back in time by simple scaling $\propto \dot{D}^{-1}$. The initial densities at $\bm{q}(\bm{x})$ can then be computed using linear theory,

$$\delta_{in} \propto -(\lambda_1 + \lambda_2 + \lambda_3) \;, \tag{37}$$

and the IPDF is computed by bin counting of $\delta_{in}$ values across the Eulerian grid, weighted by the present densities at the grid points.

A key feature of the recovered IPDF is that it is sensitive to the assumed value of $\Omega$ when the input data is velocities (§8.4), and is $\Omega$-independent if the input is density, so that it can be used to robustly recover the IPDF from the density field of the 1.2 Jy IRAS survey (Nusser *et al.* 1994). The IPDF turns out to be insensitive to galaxy biasing in the range $0.5 \leq b \leq 2$, at least for the power-law biasing relation assumed. Errors were evaluated using mock IRAS-like catalogs, and the IPDF was found *consistent with Gaussian, e.g.* the initial skewness and kurtosis are limited at the $3\sigma$ level to $-0.65 < S < 0.36$ and $-0.82 < K < 0.62$ – useful for evaluating specific non-Gaussian models. The first-year COBE measurements are consistent with Gaussian but the noise limits their discriminatory power to strongly non-Gaussian models (Smoot *et al.* 1994).

*Gaussianization.* A simple method for recovering the initial fluctuations from the galaxy distribution under the assumption that they were Gaussian is based on the assertion that both gravitational evolution and biased galaxy formation tend to preserve the rank order of density in cells. The non-Gaussian distribution of galaxies in cells simply needs to be "Gaussianized" in a rank-preserving way (Weinberg 1991). The initial conditions can then be evolved forward using an N-body code and compared with the observed galaxy distribution in $z$-space until convergence. A self-consistent solution for a redshift survey in the PP region was found if $b_o \sim 2$, while it was impossible to match the structure on small and large scales simultaneously with GI, Gaussian fluctuations and no biasing. This result derived independently of $\Omega$ implies a high $\Omega$ once $\beta_o$ is determined by another method (§8), but note that the limited surveyed region may be an "unfair" sample.

*Least action.* The general GI problem with mixed boundary conditions lends itself naturally to an application of Hamilton's action principle (Peebles 1989; 1990; 1993). The comoving orbit $\bm{x}_i(t)$ of each mass point $m_i$ is parametrized in a way that satisfies the boundary conditions

$$\bm{x}_i(t_0) = \bm{x}_{i0} \;, \quad \lim_{t \to 0} a^2 \dot{\bm{x}}_i = 0 \;, \tag{38}$$

and the action

$$S = \int_0^{t_0} L dt = \int_0^{t_0} dt \sum_i [(1/2) m_i a^2 \dot{\bm{x}}_i^2 - m_i \Phi_g(\bm{x}_i)] \tag{39}$$

is minimized to determine the free parameters. The orbits can be

$$\bm{x}_i(t) = \bm{x}_i(t_0) + \sum_n f_n(t) \bm{C}_{i,n} \;, \tag{40}$$

where $f_n(t)$ each satisfy the boundary conditions, and $\bm{C}_{i,n}$ are the parameters. The problem can be solved to any desired accuracy by increasing $n$, and a proper choice of $f_n$'s helps the series to converge rapidly to the desired solution. A generalization of the Z approximation of the sort

$$f_n(t) = [D(t) - D(t_0)]^n \qquad (41)$$

is particularly efficient (Giavalisco *et al.* 1993). In a preliminary application of this scheme to a redshift sample (Shaya *et al.* 1994) the galaxies are assumed to trace mass and be self-gravitating. The complication caused by observing redshifts is solved by an iterative procedure: a tentative guess is made for the $\boldsymbol{x}_i$ at $t_0$ based on TF distances to a sub-sample of galaxies and a crude flow model, and the least-action solution provides peculiar velocities, *i.e.* redshifts, whose deviations from the observed redshifts are used to correct the $\boldsymbol{x}_i$ for the next iteration, until convergence. After the original study of the history of the Local Group (Peebles 1989), the method was applied to a sample based on 500 groups within 30 h$^{-1}$Mpc from the Nearby Galaxy Catalog. The results indicate that the unknown tidal forces from outside the sampled volume have a significant effect on the recovered fields and should be incorporated in future applications. The solution obtained depends on the assumed $\Omega$, so a comparison with independent TF distances can in principle constrain $\Omega$ (§8.2).

## 8. THE VALUE OF $\Omega$

Assuming that the inferred motions are real and generated by GI, they can be used to estimate $\Omega$. Evidence from virialized systems on smaller scales suggest a low-density universe of $\Omega \sim 0.1 - 0.2$, but these values may be biased. The spatial *variations* of the large-scale velocity field now allow measuring the mass density in a volume closer to a "fair" sample. One family of methods is based on comparing the dynamical fields derived from velocities to the fields derived from galaxy redshifts (§9.1, 9.2). These methods can be applied in the linear regime but they always rely on the assumed biasing relation between galaxies and mass often parametrized by $b$, so they provide an estimate of $\beta \equiv f(\Omega)/b$. Another family of methods measures $\beta$ from redshift surveys alone, based on $z$-space deviations from isotropy (§9.3). Finally, there are methods which rely on non-linear effects in the velocity data alone, and provide estimates of $\Omega$ independent of $b$ (§9.4, 9.5). Note that the errors quoted by different authors reflect different degrees of sophistication in the error analysis, and are in many cases underestimates of the true uncertainty.

### 8.1 $\beta$ from Galaxies versus the CMB Dipole

Eq. (28) is best at estimating $\boldsymbol{v}(0)$, the linear velocity of the LG in the CMB frame due to the gravitational acceleration $\boldsymbol{g}(0)$ exerted by the mass fluctuations around it. A comparison with the LG velocity of $627 \pm 22$ km s$^{-1}$ as given by the CMB dipole is a direct measure of $\beta$. One way to estimate $\boldsymbol{g}(0)$ is from a whole-sky galaxy survey where only the angular positions and the fluxes (or diameters) are observed, exploiting the coincidence of nature that both the apparent flux and the gravitational force vary as $r^{-2}$. If $L \propto M$, then the vector sum of the fluxes in a volume-limited sample is $\propto \boldsymbol{g}(0)$ due to the mass in that volume. This idea can be modified to deal with a flux-limited sample once the luminosity function is known, and applications to the combined UGC/ESO diameter-limited catalog of optical galaxies yield $\beta_O$ values in the range $0.3 - 0.5$ (Lahav 1987; Lynden-Bell *et al.* 1989). These estimates suffer from limited sky coverage, uncertain corrections for Galactic extinction, and different selection procedures defining the north and south samples. The IRAS catalog provides a superior sky coverage of 96% of the sky, with negligible Galactic extinction and with fluxes observed by one telescope, but with possible under-sampling of cluster cores (Kaiser & Lahav 1989). A typical estimate from the angular IRAS catalog is $\beta_I = 0.9 \pm 0.2$ (Yahil *et al.* 1986).

The redshift surveys provide the third dimension which could help deriving $\boldsymbol{g}(0)$ by eq. (28), subject to the difficulties associated with discrete, flux-limited sampling (§5). The question is whether $\boldsymbol{g}(0)$ is indeed predominantly due to the mass within the volume sampled, *i.e.* whether $\boldsymbol{g}(0)$ as computed from successive concentric spheres converges interior to $R_{max}$. This is an issue of fundamental uncertainty (*e.g.* Lahav *et al.* 1990; Juszkiewicz *et al.* 1990; Strauss *et al.* 1992b). The $r$–$z$ mapping (21) could either compress or rarify the $z$-space volume elements depending on the sign of $u$ in the sense that an outflow makes the $z$-space density $\delta_z$ smaller than the true density $\delta$:

$$\delta_z(\boldsymbol{x}) \approx \delta(\boldsymbol{x}) - 2[\boldsymbol{v}(\boldsymbol{x}) - \boldsymbol{v}(0)] \cdot \hat{\boldsymbol{r}}/r \ . \tag{42}$$

The varying selection function adds to this geometrical effect [in analogy to $n(r)$ in the IM bias (§3.2)] and there is contribution from $d\boldsymbol{v}/dr$ as well (Kaiser 1987). It is thus clear that the redshifts must be corrected to distances and that any uncertainty in $\boldsymbol{v}(\boldsymbol{x})$ at large $\boldsymbol{x}$ or at $\boldsymbol{x}=0$ would confuse the derived $\boldsymbol{g}(0)$. The latter is the Kaiser "rocket effect": if $\boldsymbol{v}(0)$ originates from a finite volume $r < r_o$ and the density outside $r_o$ is uniform with $\boldsymbol{v} = 0$, then the measurements in $z$-space introduce a fake $\boldsymbol{g}(0)$ in the direction of $\boldsymbol{v}(0)$ due to the matter outside $r_o$, and this $\boldsymbol{g}(0)$ is logarithmically diverging with $r$. $\boldsymbol{v}(0)$ is uncertain because it is derived like the rest of $\boldsymbol{v}(\boldsymbol{x})$ from the density distribution – not from the CMB dipole. These difficulties in identifying convergence limit the effectiveness of this method in determining the PS on large scales and $\beta$. The hopes for improvement by increasing the depth are not high because the signal according to conventional PS models drops with distance faster than the shot-noise.

Attempting to measure $\beta_I$ from the IRAS data, Strauss *et al.* (1992) computed the probability distribution of $\boldsymbol{g}(0)$ under several models for the statistics of fluctuations, via a self-consistent solution for the velocities and an ad hoc fix to the rocket effect, which enabled partial corrections for shot-noise, finite volume, and small-scale non-linear effects. They confirmed that the direction of $\boldsymbol{g}(0)$ converges to a direction only $\sim 20°$ away from the CMB dipole, but were unable to determine unambiguously whether $|\boldsymbol{g}(0)|$ converges even within 100 h$^{-1}$Mpc. A maximum likelihood fit and careful error analysis constrained $\beta_I$ to the range $0.4 - 0.85$ with little sensitivity to the PS assumed. Rowan-Robinson *et al.* (1991, 1993) obtained from the QDOT dipole $\beta_I = 0.8^{+0.2}_{-0.15}$. Hudson's (1993b) best estimate from the optical dipole is $\beta_O = 0.72^{+0.37}_{-0.18}$.

The volume-limited Abell/ACO catalog of clusters with redshifts within 300 h$^{-1}$Mpc was used to compute $\boldsymbol{g}(0)$ in a similar way under the assumption that clusters trace mass linearly (Scaramella *et al.* 1991). An apparent convergence was found by $\sim 180$ h$^{-1}$Mpc to the value $\boldsymbol{g}(0) \approx 4860\beta_c$ km s$^{-1}$. A comparison with the LG-CMB motion of 600 km s$^{-1}$ yields $\beta_c \approx 0.123$, which corresponds to $\beta_O \approx 0.44$ and $\beta_I \approx 0.56$ if the ratios of biasing factors are $4.5 : 1.3 : 1$ (§7.1). A similar analysis by Plionis & Valdarnini (1991) yielded convergence by $\sim 150$ h$^{-1}$Mpc and $\beta$ values larger by $\sim 30 - 80\%$.

### 8.2 $\beta$ from Galaxy Density versus Velocities

The linear correlation found between mass density and galaxy density (§6.2) can be used to estimate the ratio $\beta$. The density $\delta_v$ determined by POTENT from velocities assuming $\Omega = 1$ relates in linear theory to the true $\delta$ by $\delta_v \propto f(\Omega)\delta$, while linear biasing assumes $\delta = b^{-1}\delta_g$, so $\delta_v = \beta\delta_g$. Dekel et al. (1993) carried out a careful likelihood analysis using the POTENT mass density from the Mark II velocity data and the density of IRAS 1.9 Jy galaxies, and found $\beta_I = 1.3^{+0.75}_{-0.6}$ at 95% confidence. A similar analysis based on the Mark III and IRAS 1.2 Jy is in progress. The degeneracy of $\Omega$ and $b$ is broken in the quasi-linear regime, where $\delta(\boldsymbol{v})$ is no longer $\propto f^{-1}$. The compatible quasi-linear corrections in POTENT and in the IRAS analysis allow a preliminary attempt to separate these parameters, which yields for Mark II data $\Omega > 0.46$ (95% level) if $b_I > 0.5$. A correction for IM bias could reduce the 95% confidence limit to $\Omega > 0.3$ at most. These results are valid for linear biasing; possible non-linear biasing may complicate the analysis because it is hard to distinguish from non-linear gravitational effects.

The advantage of comparing densities is that they are *local*, independent of reference frame, and can be reasonably corrected for non-linear effects. The comparison can alternatively be done between the observed velocities and those predicted from a redshift survey, subject to limited knowledge of the quadrupole and higher moments of the mass distribution outside the surveyed volume and other biases. Kaiser et al. (1991) obtained from Mark II velocities versus QDOT predictions $\beta_I = 0.9^{+0.20}_{-0.15}$. An analysis by Roth (1994) using IRAS 1.9 Jy galaxies yielded $\beta_I = 0.6 \pm 0.3$ ($2\sigma$). Nusser & Davis (1994) implemented a novel method based on the Z approximation in spherical harmonics to derive the velocity dipole of distant shells from the IRAS 1.2 Jy survey and found in comparison to the dipole of observed velocities $\beta_I = 0.6 \pm 0.2$.

Similar comparisons with the optical galaxy fields indicate a similar correlation between light and mass. A comparison at the velocity level gives (Hudson 1994) $\beta_O = 0.5 \pm 0.1$, and a preliminary comparison at the density level with 12 $h^{-1}$Mpc smoothing indicates (Hudson et al. 1994) $\beta_O \approx 0.6 \pm 0.2$, in general agreement with the ratio of $b_O/b_I \approx 1.3$–$1.4$ obtained by direct comparison. Shaya et al. (1994) applied the least-action reconstruction method (§7.2) to a redshift survey of several hundred spirals within our local 30 $h^{-1}$Mpc neighborhood and crudely obtained by comparison to TF distances $\beta_O \sim 0.4$.

### 8.3 $\beta$ from Distortions in Redshift Space

Redshift samples, which contain hidden information about velocities, can be used on their own to measure $\beta$. The clustering, assumed isotropic in real space, $\boldsymbol{x}$, is anisotropic in $z$-space, $\boldsymbol{z}$, where $z = r + \hat{\boldsymbol{x}} \cdot \boldsymbol{v}$ displaces galaxies along the preferred direction $\hat{\boldsymbol{x}}$. While virial velocities on small scales stretch clusters into "fingers of god" along the LOS, systematic infall motions enhance large-scale structures by artificially squashing them along the LOS. The linear approximation $-\boldsymbol{\nabla}\cdot\boldsymbol{v} = \beta\delta_g$ indicates that the effect is $\beta$ dependent because $-\boldsymbol{\nabla}\cdot\boldsymbol{v}$ is related to the anisotropy in $z$-space while $\delta_g$ is isotropic, so the statistical deviations from isotropy can tell $\beta$ (e.g. Sargent & Turner 1977).

Kaiser (1987) showed in linear theory that the anisotropic Fourier PS in $z$-space is related to the $r$-space PS of mass density, $P(k)$, via

$$P^z(k,\mu) = P(k)(1 + \beta\mu^2)^2 \; , \tag{43}$$

where $\mu \equiv \hat{\boldsymbol{k}} \cdot \hat{\boldsymbol{x}}$. This relation is valid only for a fixed $\mu$, *i.e.* in a distant volume of small solid angle (Zaroubi & Hoffman 1994), but there are ways to apply it more generally. The redshift PS can be decomposed into Legendre polynomials, $\mathcal{P}_l(\mu)$, with even multipole moments $P_l^z(k)$,

$$P^z(k,\mu) = \sum_{l=0}^{\infty} P_l^z(k)\mathcal{P}_l(\mu) \,, \quad P_l^z(k) = \frac{2l+1}{2} \int_{-1}^{+1} d\mu P^z(k,\mu)\mathcal{P}_l(\mu) \,. \tag{44}$$

Based on eq. (43) the first two non-vanishing moments are

$$P_0^z(k) = (1 + \frac{2}{3}\beta + \frac{1}{5}\beta^2)P(k) \,, \quad P_2^z(k) = (\frac{4}{3}\beta + \frac{4}{7}\beta^2)P(k) \,, \tag{45}$$

so the observable ratio of quadrupole to monopole is a function of $\beta$ independent of $P(k)$. A preliminary application to the 1.2 Jy IRAS survey yields $\beta_I \sim 0.3 - 0.4$ at wavelength $30 - 40$ $h^{-1}$Mpc, suspected of being an underestimate because of non-linear effects out to $\sim 50$ $h^{-1}$Mpc (Cole et al. 1993). Peacock & Dodds (1994) developed a method for reconstructing the linear PS and they obtain $\beta_I = 1.0 \pm 0.2$.

The distortions should be apparent in the $z$-space two-point correlation function, $\xi_z(r_p,\pi)$, the excess of pairs with separation $\pi$ along the LOS and $r_p$ transversely (Davis & Peebles 1983). The contours of equal $\xi$, assumed round in $r$-space, appear in $z$-space elongated along the LOS at small separations and squashed on large scales depending on $\beta$. Hamilton (1992; 1993) used the multiple moments of $\xi_z$, in analogy to eqs. (26-27), and his various estimates from the 1.9 Jy IRAS survey span the range $\beta_I = 0.25 - 1$. Fisher et al. (1994a) computed $\xi^z(r_p,\pi)$ from the 1.2 Jy IRAS survey, and derived the first two pair-velocity moments. Their attempt to use the velocity dispersion via the Cosmic Virial Theorem led to the conclusion that this is a bad method for estimating $\Omega$, but the mean, $\langle v_{12} \rangle = 109^{+64}_{-47}$ at 10 $h^{-1}$Mpc, yielded $\beta_I = 0.45^{+0.27}_{-0.18}$. The drawbacks of using $\xi$ versus PS are that (a) the uncertainty in the mean density affects all scales in $\xi$ whereas it is limited to the $k = 0$ mode of the PS, (b) the errors on different scales in $\xi$ are correlated whereas they are independent in a linear PS for a Gaussian field, and (c) $\xi$ mixes different physical scales, complicating the transition between the linear and non-linear regimes. Non-linear effects tend to make all the above results underestimates.

A promising method that is tailored to deal with a realistic redshift survey of a selection function $\phi(r)$ and does not rely on the subtleties of eq. (43) is based on a weighted spherical harmonic decomposition of $\delta_z(\boldsymbol{z})$ (Fisher et al. 1994b),

$$a_{lm}^z = \int d^3z \, \phi(r) f(z) [1 + \delta_z(\boldsymbol{z})] Y_{lm}(\hat{\boldsymbol{s}}), \quad \langle |a_{lm}^z|^2 \rangle = \frac{2}{\pi} \int_0^\infty dk \, k^2 P(k) |\psi_l^r(k) + \beta \psi_l^c(k)|^2 \,. \tag{46}$$

The arbitrary weighting function $f(z)$ is vanishing at infinity to eliminate surface terms. The mean-square of the harmonics is derived in linear theory assuming that the survey is a "fair" sample, and $\psi^r$ and $\psi^c$ are explicit integrals over $r$ of certain expressions involving $\phi(r)$, $f(r)$, Bessel functions and their derivatives. The first term represents real structure and the second is the correction embodying the $z$-space distortions. The harmonic PS in $z$-space, averaged over $m$, is thus determined by $P(k)$ and $\beta$, where the $z$-space distortions appear as a $\beta$-dependent excess at small $l$. The harmonic PS derived from the 1.2 Jy IRAS survey yield $\beta_I = 1.0 \pm 0.3$ for assumed $\sigma_8 = 0.7$ (motivated by the IRAS $\xi$, Fisher *et al.* 1994a), with an additional systematic uncertainty of $\pm 0.2$ arising from the unknown shape of the PS.

The methods for measuring $\beta$ from redshift distortions are promising because they are relatively free of systematic errors and because very large redshift surveys are achievable in the near future. With a sufficiently large redshift survey, one can even hope to be able to use the non-linear effects to determine $\Omega$ and $b$ separately.

### 8.4 $\Omega$ from PDFs using velocities

Assuming that the initial fluctuations are a random *Gaussian* field, the one-point PDF of smoothed density develops a characteristic skewness due to non-linear effects early in the quasi-linear regime (§7.2). The skewness of $\delta$ is given in second-order perturbation theory by

$$\langle \delta^3 \rangle / \langle \delta^2 \rangle^2 = (34/7 - 3 - n) , \qquad (47)$$

with $n$ the effective power index near the smoothing scale (Bouchet *et al.* 1992). Since this ratio for $\delta$ is practically independent of $\Omega$, and since $\boldsymbol{\nabla} \cdot \boldsymbol{v} \sim -f\delta$, the corresponding ratio for $\boldsymbol{\nabla} \cdot \boldsymbol{v}$ strongly depends on $\Omega$, and in second-order (Bernardeau *et al.* 1994)

$$S_3 \equiv \langle (\boldsymbol{\nabla} \cdot \boldsymbol{v})^3 \rangle / \langle (\boldsymbol{\nabla} \cdot \boldsymbol{v})^2 \rangle^2 = -f(\Omega)^{-1}(26/7 - 3 - n) . \qquad (48)$$

Using N-body simulations and 12 h$^{-1}$Mpc smoothing one indeed finds $S_3 = -1.8 \pm 0.7$ for $\Omega = 1$ and $S_3 = -4.1 \pm 1.3$ for $\Omega = 0.3$, where the quoted error is the cosmic scatter for a sphere of radius 40 h$^{-1}$Mpc in a CDM universe ($H_0 = 75$, $b = 1$). A preliminary estimate of $S_3$ in the current POTENT velocity field within 40 h$^{-1}$Mpc is $-1.1 \pm 0.8$, the error representing distance errors. With the two errors added in quadrature, $\Omega = 0.3$ is rejected at the $\sim 2\sigma$ level (somewhat sensitive to the assumed PS).

Since the PDF contains only part of the information stored in the data and is in some cases not that sensitive to the IPDF (§7.2), a more powerful bound can be obtained by using the detailed $\boldsymbol{v}(\boldsymbol{x})$ to recover the IPDF, and use the latter to constrain $\Omega$. This is done by comparing the $\Omega$-dependent IPDF recovered from observed velocities to an assumed IPDF(Nusser & Dekel 1993), most naturally a Gaussian as recovered from IRAS density (§7.2). The velocity out of POTENT Mark II within a conservatively selected volume was fed into the IPDF recovery procedure with $\Omega$ either 1 or 0.3, and the errors due to distance errors and cosmic scatter were estimated. The IPDF recovered with $\Omega = 1$ is found marginally consistent with Gaussian while the one recovered with $\Omega = 0.3$ shows significant deviations. The largest deviation bin by bin in the IPDF is $\sim 2\sigma$ for $\Omega = 1$ and $> 4\sigma$ for $\Omega = 0.3$, and a similar rejection is obtained with a $\chi^2$-type statistic. The skewness and kurtosis are poorly determined because of noisy tails but the replacements $\langle x|x| \rangle$ and $\langle |x| \rangle$ allow a rejection of $\Omega = 0.3$ at the $(5-6)\sigma$ levels.

## 8.5 Ω from Velocities in Voids

A diverging flow in an extended low-density region can provide a robust dynamical lower bound on Ω, based on the fact that large outflows are not expected in a low-Ω universe (Dekel & Rees 1994). The velocities are assumed to be induced by GI, but *no* assumptions need to be made regarding galaxy *biasing* or the exact statistical nature of the fluctuations. The derivatives of a diverging velocity field infer a non-linear approximation to the mass density, $\delta_c(\Omega, \partial \boldsymbol{v}/\partial \boldsymbol{x})$ (eq. 7), which is an overestimate, $\delta_c > \delta$, when the true value of Ω is assumed. Analogously to $\delta_0 = -f(\Omega)^{-1} \boldsymbol{\nabla} \cdot \boldsymbol{v}$, the $\delta_c$ inferred from a given diverging velocity field becomes more negative when a smaller Ω is assumed, and it may become smaller than $-1$. Since $\delta \geq -1$ because mass is never negative, Ω is bounded from below.

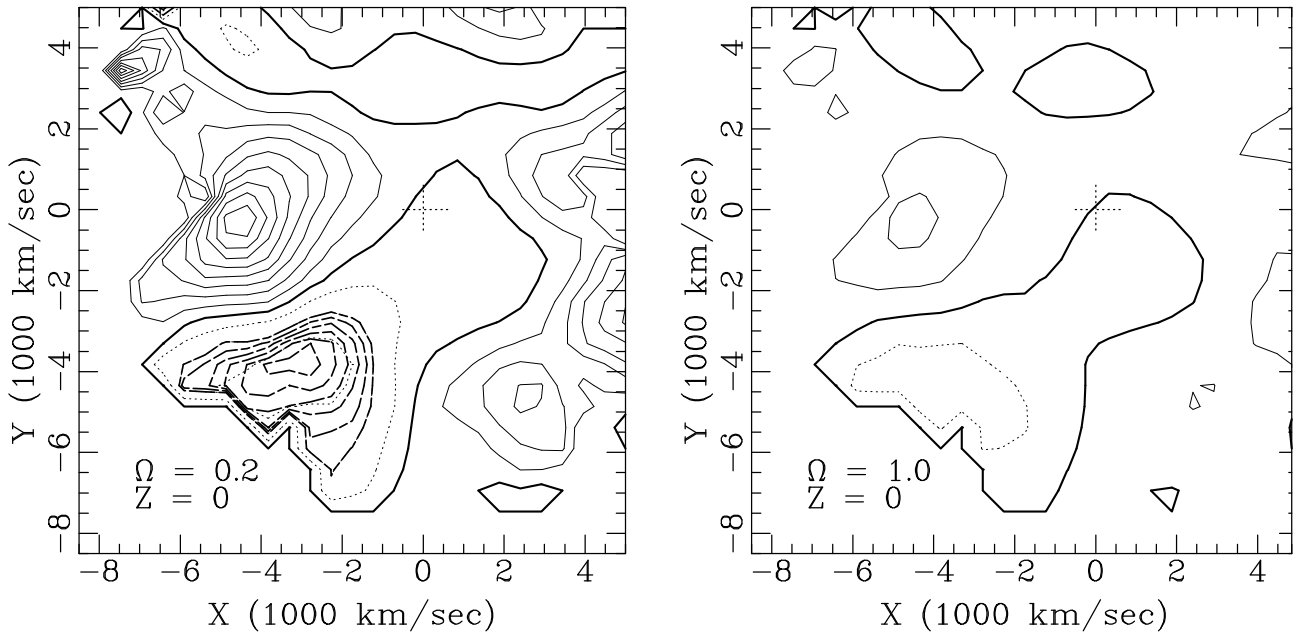

**Figure 8:** Maps of $\delta_c$ inferred from the observed velocities near the Sculptor void in the Supergalactic plane, for two values of Ω. The LG is marked by '+' and the void is confined by the Pavo part of the GA (left) and the Aquarius extension of PP (right). Contour spacing is 0.5, with $\delta_c = 0$ heavy, $\delta_c > 0$ solid, and $\delta_c < 0$ dotted. The heavy-dashed contours mark the illegitimate downward deviation of $\delta_c$ below $-1$ in units of $\sigma_\delta$, starting from zero (*i.e.* $\delta_c = -1$), and decreasing with spacing $-0.5\sigma$. The value $\Omega = 0.2$ is ruled out at the $2.9\sigma$ level (Dekel & Rees 1994).

The inferred $\delta_c(\boldsymbol{x})$ smoothed at $12\,h^{-1}\mathrm{Mpc}$ and the associated error field $\sigma_\delta$ are derived by POTENT from the observed radial velocities and, focusing on the deepest density wells, the assumed Ω is lowered until $\delta_c$ becomes significantly smaller than $-1$. The most promising "test case" provided by the Mark III data seems to be a broad diverging region centered near the supergalactic plane at the vicinity of $(X, Y) = (-25, -40)$ in $h^{-1}\mathrm{Mpc}$ – the "Sculptor void" of galaxies (Kauffman *et al.* 1991) next to the "Southern Wall" (Figure 8). Values of $\Omega \approx 1$ are perfectly consistent with the data, but $\delta_c$ becomes smaller than $-1$ already for $\Omega = 0.6$. The values $\Omega = 0.3$ and $0.2$ are ruled out at the 2.4-, and $2.9\sigma$ levels in terms of the random error $\sigma_\delta$. This is just a preliminary result. The systematic errors have been partially corrected for in POTENT, but a more specific investigation of the SG biases affecting the smoothed velocity field in density wells is required. For the method to be effective one needs to find a void that is (a) bigger than the correlation length for its vicinity to represent the universal $\Omega$, (b) deep enough for the lower bound to be tight, (c) nearby enough for the distance errors to be small, and (d) properly sampled to trace the velocity field in its vicinity.

### TABLE 1: $\Omega$ and $b$[†]

| | | | |
|---|---|---|---|
| CMB dipole | vs galaxies angular | Yahil *et al.* 86 | $\beta_I = 0.9 \pm 0.2$[‡] |
| | vs galaxies redshift | Strauss *et al.* 92 | $\beta_I = 0.4 - 0.85$ |
| | | Rowan-Rob. *et al.* 91 | $\beta_I = 0.8^{+0.2}_{-0.15}$ |
| | vs galaxies angular | Lynden-Bell *et al.* 89 | $\beta_O = 0.3 - 0.5$ |
| | vs galaxies redshift | Hudson 93b | $\beta_O = 0.7^{+0.4}_{-0.2}$ |
| | clusters | Scaramella *et al.* 91 | $\beta_C \sim 0.13$ |
| | | Plionis *et al.* 91 | $\beta_C \sim 0.17 - 0.22$ |
| $\boldsymbol{v}$ vs $\delta_g$ | Potent-IRAS1.9 density | Dekel *et al.* 93 | $\beta_I = 1.3^{+0.75}_{-0.6}$ (95%) |
| | Potent-IRAS1.2 v-dipole | Nusser & Davis 94 | $\beta_I = 0.6 \pm 0.2$ |
| | TF-QDOT | Kaiser *et al.* 91 | $\beta_I = 0.9^{+0.2}_{-0.15}$ |
| | TF-QDOT clusters | Frenk *et al.* 94 | $\beta_I = 1.0 \pm 0.3$ |
| | TF inverse - IRAS1.9 | Roth 94 | $\beta_I = 0.6 \pm 0.35$ ($2\sigma$) |
| | Potent-Optical density | Hudson *et al.* 94 | $\beta_O = 0.8 \pm 0.2$ |
| | TF-Optical | Hudson 93c | $\beta_O = 0.5 \pm 0.1$ |
| | TF-Optical local | Shaya *et al.* 94 | $\beta_O \sim 0.4$ |
| $z$-distortions | $P_k$ IRAS1.2 | Cole *et al.* 93 | $\beta_I \gtrsim 0.3 - 0.4$ |
| | $P_k$ IRAS1.2 | Peacock & Dodds 94 | $\beta_I = 1.0 \pm 0.2$ |
| | $\xi$ IRAS1.9 | Hamilton 93 | $\beta_I \sim 0.25 - 1$ |
| | $\xi$ IRAS1.2 | Fisher *et al.* 94a | $\beta_I = 0.45^{+0.3}_{-0.2}$ |
| | $Y_{lm}$ IRAS1.2 | Fisher *et al.* 94b | $\beta_I = 1.0 \pm 0.3$ |
| Velocities | Gaussian IPDF Potent | Nusser & Dekel 93 | $\Omega > 0.3$ ($4\sigma - 6\sigma$) |
| | Skew($\boldsymbol{\nabla} \cdot \boldsymbol{v}$) Potent | Bernardeau *et al.* 94 | $\Omega > 0.3$ ($2\sigma$) |
| | Voids Potent | Dekel & Rees 94 | $\Omega > 0.3$ ($2.4\sigma$) |

[†]$\beta \equiv \Omega^{0.6}/b$, $b_C : b_O : b_I \approx 4.5 : 1.3 : 1.0$, see text.
[‡]All errors are $1\sigma$ unless stated otherwise.

# 9. DISCUSSION: ARE THE HYPOTHESES JUSTIFIED?

*Are the motions real?* The large-scale bulk flow rests upon the interpretation of the CMB dipole as motion of the LG in the CMB frame, based on the *assumption* of cosmological isotropy (H1), and supported by the gravitational acceleration derived at the LG from the galaxy and mass distribution around it (§6.1). The detection of $\delta T/T \sim 10^{-5}$ is a clear indication, via GI, for $\sim 300$ km s$^{-1}$ motions over $\sim 100$ h$^{-1}$Mpc (§6.1). While this is reassuring, the LP hint for possible very-large coherence suggests caution (§7.1). The evidence in support of H6, that the TF-inferred motions about the LG are real, are (a) the correlation $\delta_g \propto -\boldsymbol{\nabla} \cdot \boldsymbol{v}$, which is robustly predicted for true velocities based on continuity and would be hard to mimic by environmental effects (§6.2), (b) the failure to detect any significant correlation between velocities and the environment or other galaxy properties (§6.3), and (c) the similarity between the velocity fields traced by spirals (TF) and by ellipticals ($D_n$–$\sigma$) (§6.3).

Is *linear biasing* a good approximation (H5)? The galaxy-velocity correlation is most sensitive to it, and the observed correlation on scales $\gtrsim 10$ h$^{-1}$Mpc is consistent with linear biasing properly modified in the tails (§6.2). However, it is difficult to distinguish non-linear biasing from non-linear gravitational effects, and the range of different estimates of $\beta$ (§8) may indicate that the biasing parameter varies as a function of scale. The ratio of $\sim 10$ h$^{-1}$Mpc-smoothed densities for optical and IRAS galaxies is $b_o/b_I \approx 1.3 - 1.5$.

Is *gravity* the dominant source of LSS (H4)? The observed velocity-density correlation (§6.2) is fully consistent with GI, but it is sensitive to continuity more than to the specific time dependence implied by gravity. Any non-GI process followed by a gravitating phase would end up consistent with this observation, and certain non-GI models may show a similar spatial behavior even if gravity never plays any role. The E-S correlation (§6.3) is also consistent with gravity as galaxies of all types trace the same velocity field (H4b), but any model where all galaxies are set into motion by the same mechanism could pass this test. The strongest evidence for gravitational origin comes from the statistical agreement between the fluctuations of today and those implied by the CMB at the time of recombination. A marginal warning signal for GI is provided by the $\sim 700$ km s$^{-1}$ bulk velocity indicated by LP for rich clusters across $\sim 200$ h$^{-1}$Mpc. Such a velocity at face value would be in conflict with the gravitational acceleration implied by the cluster distribution and with the $\delta T/T \sim 10^{-5}$ at $\sim 2°$, but the errors are big.

The property of *irrotationality* (H4a) used in the reconstruction from either velocities or densities is impossible to deduce solely from observations of velocities along the lines of sight from one origin. Irrotationality is assumed based on the theory of GI, or it can be tested against the assumption of isotropy by measuring the isotropy of the velocity field derived by potential analysis for a fair sample.

The observed CMB fluctuations provide evidence for *initial fluctuations* (H2), consistent with a scale-invariant $n \sim 1$ spectrum. The observed motions are also consistent with $n \sim 1$ (§7.1) (with the uncertain LP result as a possible exception). The indications for somewhat higher large-scale power in the clustering of galaxies (Maddox *et al.* 1990) may reflect non-trivial biasing. The question of whether the fluctuations were *Gaussian* (H2) cannot be answered by observed velocities alone. The PDF of $\boldsymbol{\nabla} \cdot \boldsymbol{v}$ is consistent with Gaussian initial fluctuations skewed by non-linear gravity, but this is not a very discriminatory test. Nevertheless, the galaxy spatial distribution does indicate Gaussian initial fluctuations fairly convincingly (§7.2)

Can we tell the nature of the *dark matter* (H3)? In view of the tight nucleosynthesis constraints on baryonic density, the high $\Omega$ indicated by the motions requires non-baryonic DM. The mass-density PS on scales $10-100$ h$^{-1}$Mpc is calculable in principle but the current uncertainties do not allow a clear distinction between the possibilities of baryonic, cold, hot or mixed DM. The mixed model seems to score best in view of the overall LSS data, as expected from a model with more free parameters, but CDM in fact does somewhat better in fitting the large-scale motions. I do not think that any of the front-runner models is significantly ruled out at this point, contrary to occasional premature statements in the literature about the "death" of certain models. I predict that were the DM constituent(s) to be securely detected in the lab, the corresponding scenario of LSS would find a way to overcome the $\sim 2\sigma$ obstacles it is facing now.

What can we conclude about the *background cosmology* (H1)? All the observations so far are consistent with large-scale homogeneity and isotropy (with the exception of the $2\sigma$ LP discrepancy). The motions say nothing about $H$ or $\Lambda$ (Lahav *et al.* 1991), but they provide a unique opportunity to constrain $\Omega$ in several different ways. Some methods put a strong ($>3\sigma$) lower bound of $\Omega > 0.2-0.3$. This is consistent with the theoretically-favored $\Omega = 1$ but "ugly" values near $\Omega \approx 0.5$ are not ruled out either. The apparent range of $\beta$ values obtained on different scales is partly due to errors, and any remaining difference may be explained by a scale-dependent non-linear biasing relation between the different galaxy types and mass. The data are thus at least consistent with the predictions of *Inflation*: flat geometry and Gaussian, scale-invariant initial fluctuations. Recall however that $\Omega = 1$ predicts $t_0 = 6.3h^{-1}$Gyr, which will be in conflict with the age constraints from globular clusters, $t_0 = 15 \pm 3$Gyr, if the Hubble constant $h$ is not close to 0.5.

The rapid progress in this field guarantees that many of the results and uncertainties discussed above will soon become obsolete, but I hope that the discussion of concepts will be of lasting value, and that the methods discussed can be useful as are and as a basis for improvements.

I thank G. Ganon, Y. Hoffman, T. Kolatt, S. Markoff and I. Zehavi for assistance, M. Hudson and A. Yahil for plots, and G. Blumenthal, S.M. Faber, O. Lahav, M. Strauss, D. Weinberg, i J. Willick, and A. Yahil for very helpful comments. This work has been supported by grants from the US-Israel Binational Science Foundation and the Israel Basic Research Foundation.

# REFERENCES

Aaronson, M., Bothun, G.D., Cornell, M.E., Dawe, J.A., Dickens, R.J., *et al.* 1989, *Ap. J.* **338**, 654.

Aaronson, M., Bothun, G., Mould, J., Huchra, J. Schommer, R.A., & Cornell, M.E. 1986, *Ap. J.* **302**, 536

Aaronson, M., Huchra, J., & Mould, J. 1979, *Ap. J.* **229**, 1

Aaronson, M., Huchra, J., Mould, J., Schechter, P.L., & Tully, R.B. 1982b, *Ap. J.* **258**, 64

Aaronson, M., Huchra, J., Mould, J.R., Tully, R.B., Fisher, J.R., *et al.* 1982a, *Ap. J. Supp.* **50**, 241.

Aaronson, M., & Mould, J. 1983, *Ap. J.* **265**, 1-17

Babul, A., Weinberg, D., Dekel, A., & Ostriker, J.P. 1994, *Ap. J.*, in press

Bardeen, J., Bond, J.R., Kaiser, N. & Szalay, A. 1986, *Ap. J.* **304**, 15

Bernardeau, F. 1992, *Ap. J. Lett* **390**, L61

Bernardeau, F., Juszkiewicz, R., Bouchet, F., & Dekel, A. 1994, in preparation

Bertschinger, E. 1991, in *Physical Cosmology*, ed. M. Lachieze-Rey (Editor Frontier)

Bertschinger, E., & Dekel, A. 1989, *Ap. J. Lett* **336**, L5

Bertschinger, E., Dekel, A., Faber, S.M., Dressler, A., & Burstein, D. 1990, *Ap. J.* **364**, 370

Bertschinger, E., Gorski, K., & Dekel, A. 1990, *Nature* **345**, 507

Bertschinger, E., & Juszkiewicz, R. 1988, *Ap. J. Lett* **334**, L59

Blumenthal, G.R., Dekel, A., & Primack, J.R. 1988, *Ap. J.* **326**, 539

Bothun, G. D., Aaronson, M., Schommer, B., Huchra, J., & Mould, J. 1984, *Ap. J.* **278**, 475

Bouchet, F., Juszkiewicz, R., Colombi, S., & Pellat, R. 1992, *Ap. J. Lett* **394**, L5

Bouchet, F., Strauss, M., Davis, M., Fisher, K.B., Yahil, A., & Huchra, J.P. 1993, *Ap. J.* **417**, 36

Burstein, D. 1990a, in *Large Scale Structure and Peculiar Motions in the Universe*, eds. D. W. Latham & L. N. Da Costa (ASP Conference Series)

Burstein, D. 1990b, *Rep. Prog. Phys.* **53**, 421-81

Burstein, D., Davies, R.L., Dressler, A., Faber, S.M., Lynden-Bell, D., Terlevich, R.J., & Wegner, G. 1986 in *Galaxy Distances and Deviations from Universal Expansion*, eds. B.F. Madore & R.B. Tully (Dordrecht: Reidel) p 123-30

Burstein, D., Faber, S.M., & Dressler, A. 1990 *Ap. J.* **354**, 18

Cen, R., & Ostriker, J.P. 1993, *Ap. J.*, in press

Cheng, E.S. *et al.* 1993, *Ap. J. Lett*, in press.

Cole, S., Fisher, K.B., & Weinberg, D. 1993, *MNRAS*, in press

Colless *et al.* 1993, *MNRAS* **262**, 475

Corey, B.E., & Wilkinson, D.T. 1976, *Bull. Am. Aston. Soc.* **8**, 35

Courteau, S. 1992, Ph.D. Thesis, UCSC

Courteau, S., Faber, S.M., Dressler, A., & Willick J.A. 1993, *Ap. J. Lett* **412**, L51

Davis, M., Summers, F.J., & Schlegel D. 1992, *Nature* **359**, 393

Davis, M., & Peebles, P.J.E. 1983, *Annu. Rev. Astron. Astrophys.* **21**, 109-30

Dekel, A. 1981, *Astron. Astrophys.* **101**, 79-87

Dekel, A., Bertschinger, E., & Faber, S.M. 1990, *Ap. J.* **364**, 349

Dekel, A., Bertschinger, E., Yahil, A., Strauss, M., Davis, M., & Huchra, J. 1993, *Ap. J.* **412**, 1

Dekel *et al.* 1994, *Ap. J.* , in preparation

Dekel, A. & Rees, M.J. 1987, *Nature* **326**, 455

Dekel, A. & Rees, M.J. 1994, *Ap. J. Lett* , in press

Dekel, A., & Silk, J. 1986, *Ap. J.* **303**, 39

de Lapparent, V., Geller, M.J. Huchra, J.P. 1986, *Ap. J. Lett* **302**, L1

Djorgovski, S., & Davis, M. 1987, *Ap. J.* **313**, 59

Djorgovski, S., de Carvalho, R., & Han, M.S. 1989, in *The Extragalactic Distance Scale*, ed. S. van den Bergh & C. J. Pritchet (Provo: ASP), p 3

Dressler, A., in *Cosmic Velocity Fields* (IAP, Paris), ed. F. Bouchet & M. Lachieze-Rey , in press

Dressler, A., & Faber, S.M. 1991, *Ap. J.* **368**, 54

Dressler, A., Lynden-Bell, D., Burstein, D., Davies, R. L., Faber, S. M., Terlevich, & Wegner, G. 1987, *Ap. J.* **313**, 42

Efstathious, G. 1990, in *Physics of the Early Universe*, eds. J.A. Peacock, A.F. Heavens, & A.T. Davies (Edinburgh: SUSSP)

Faber, S. M., & Burstein, D. 1988, in the Vatican Study Week on Large Scale Motions in the Universe, eds. G. V. Coyne & V. C. Rubin (Princeton: Princeton University Press) p. 116

Faber *et al.* 1994, *Ap. J.* , in preparation.

Faber, S.M., & Jackson, R.E. 1976, *Ap. J.* **204**, 668-83

Faber, S. M., Wegner, G., Burstein, D., Davies, R. L., Dressler, A., Lynden-Bell, D., & Terlevich, R. J. 1989, *Ap. J. Supp.* **69**, 763

Feldman, H.A., & Watkins, R. 1994, in *Cosmic Velocity Fields* (IAP, Paris), ed. F. Bouchet & M. Lachieze-Rey , in press

Fisher, K.B. 1992, Ph.D. Thesis, UCB.

Fisher, K.B., Davis, M., Strauss, M.A., Yahil, A., & Huchra, J.P. 1994a, *MNRAS* , in press.

Fisher, K.B., Scharf, C.A., & Lahav, O. 1994b, *MNRAS* , in press.

Frenk, C., Kaiser, N., & Lucey, J. 1994, in preparation.

Giavalisco, M., Mancinelli, B., Mancinelli, P.J., & Yahil, A. 1993, *Ap. J.* **411**, 9

Gramman, M. 1993a, *Ap. J. Lett* **405**, L47

Gramman, M. 1993b, *Ap. J.* **405**, 449

Gregg, M.J., 1993, *Ap. J.* , ?

Gull, S., & Daniell, X. 1978, *Nature* **272**, 686

Gunn, J.E. 1989, in *the Extragalactic Distance Scale*, ed S. van den Berg & C.J. Pritchet (Provo: ASP), p. 344

Hamilton, A. J. S. 1992, *Ap. J. Lett* **385**, L5

Hamilton, A. J. S. 1993, *Ap. J. Lett* **406**, L47

Han, M.S. & Mould, J.R. 1990, *Ap. J.* **360**, 448

Han, M.S. & Mould, J.R. 1992, *Ap. J.* **396**, 453

Hudson, M. 1993a; *MNRAS* **265**, 43

Hudson, M. 1993b; *MNRAS* **265**, 72

Hudson, M. 1994; *MNRAS* **266**, 475

Hudson, M. *et al.* 1994, in preparation

Hoffman, Y. 1994, in *Cosmic Velocity Fields* (IAP, Paris), ed. F. Bouchet & M. Lachieze-Rey , in press

Hoffman, Y., & Ribak, E. 1991, *Ap. J. Lett* **380**, L5

Hogan, C.J., & White, S.D.M. 1986, *Nature* **321**, 575

Ikeuchi, S. 1981, *Pub. Astr. Soc. Japan.* **33**, 211

Jacoby, G.H., et al. 1992, *Publ. Astron. Soc. Pac.* **104**, 599-662

Juszkiewicz, R., Vittorio, N., & Wyse, R.F.G. 1990, *Ap. J.* **349**, 408

Kaiser, N. 1987, *MNRAS* **227**, 1-21

Kaiser, N. 1988, *MNRAS* **231**, 149-68

Kaiser, N. 1984, *Ap. J. Lett* **284**, L9

Kaiser,N., Efstathiou, G., Ellis, R., Frenk, C., Lawrence, A., Rowan-Robinson, M., & Saunders, W. 1991, *MNRAS* **252**, 1

Kaiser, N., & Lahav, O. 1989, *MNRAS* **237**, 129

Kaiser, N., & Stebbins, A. 1991, in *Large Scale Structure and Peculiar Motions in the Universe*, eds. D. W. Latham & L. N. Da Costa (ASP Conference Series) p. 111

Kauffman, G., & Fairall, A.P. 1991, *MNRAS* **248**, 313

Klypin, A., Holtzman, J., Primack, J.R., & Regos, E. 1993, *Ap. J.* **416**, 1

Kofman, L., Bertschinger, E., Gelb, J., Nusser, A., & Dekel, A. 1994, *Ap. J.* **420**, 44-57

Kofman, L., Blumenthal, G.R., Hodges, H., & Primack, J.R. 1990, in *Large Scale Structure and Peculiar Motions in the Universe*, eds. D. W. Latham & L. N. Da Costa (ASP Conference Series)

Kogut, A. *et al.* 1993, *Ap. J.* **419**, 1-6

Kolatt, T. & Dekel, A. 1994a, *Ap. J.* , in press

Kolatt, T. & Dekel, A. 1994b, *Ap. J.* , submitted

Kolatt, T., Dekel, A., & Lahav, O. 1994, *MNRAS* , submitted

Kolb, E.W., & Turner, M.S. 1990 *The Early Universe* (Eddison-Wesley).

Lahav, O. 1987, *MNRAS* **225**, 213-20

Lahav, O., Fisher, K.B., Hoffman, Y., Scharf, C.A., & Zaroubi, S. 1994, *Ap. J. Lett* , in press

Lahav, O., Kaiser, N., & Hoffman, Y. 1990, *Ap. J.* **352**, 448

Lahav, O., Lilje, P.B., Primack, J.R., & Rees, M.J. 1991, *MNRAS* **251**, 128

Landy, S., & Szalay, A. 1992, *Ap. J.* **391**, 494

Lauer, T.R., & Postman, M. 1993, *Ap. J.* , in press

Lucey, J.R., & Carter, D. 1988, *MNRAS* **235**, 1177

Lynden Bell, D., Faber, S.M., Burstein, D., Davies, R.L., Dressler, A., Terlevich, R. J., & Wegner, G. 1988, *Ap. J.* **326**, 19

Lynden-Bell, D., Lahav, O., & Burstein, D. 1989, *MNRAS* **241**, 325-45

Maddox, S.J, Efstathiou, G., Sutherland, W.J., & Loveday, J. 1990, *MNRAS* **242**, 43p

Mancinelli, P.J., Yahil, A., Ganon, G., & Dekel, A. 1994, in *Cosmic Velocity Fields* (IAP, Paris), ed. F. Bouchet & M. Lachieze-Rey , in press

Mathewson, D.S., Ford, V.L., & Buchhorn, M. 1992, *Ap. J. Supp.* **81**, 413

Mould, J.R. *et al.* 1991, *Ap. J.* **383**, 467

Moutarde, F., Alimi, J.-M., Bouchet. F. R., Pellat, R., & Ramani, A. 1991, *Ap. J.* **382**, 377

Nusser, A. & Dekel, A. 1992, *Ap. J.* **391**, 443

Nusser, A. & Dekel, A. 1993, *Ap. J.* **405**, 437

Nusser, A., Dekel, A., Bertschinger, E., & Blumenthal, G.R., 1991, *Ap. J.* **379**, 6

Nusser, A., Dekel, A., & Yahil, A. 1994, *Ap. J.* , in press

Nusser, A., & Davis, M. 1994, *Ap. J. Lett* , in press

Ostriker, J.P., & Cowie, L.L 1981, *Ap. J. Lett* **243**, L127

Ostriker, J.P., & Suto, Y. 1990, *Ap. J.* **348**, 378

Paczynski, B. & Piran, T. 1990, *Ap. J.* **364**, 341

Peacock, J.A., & Dodds, S.J. 1994, *MNRAS* , in press

Peebles, P.J.E. 1980, *The Large-Scale Structure of the Universe* (Princeton: Princeton University Press)

Peebles, P.J.E. 1989, *Ap. J. Lett* **344**, L53

Peebles, P.J.E. 1990, *Ap. J.* **362**, 1

Peebles, P.J.E. 1993, *Principles of Physical Cosmology* (Princeton: Princeton University Press)

Plionis, M. & Valdarnini, R. 1991, *MNRAS* **249**, 46

Rauzy, S., Lachieze-Rey, M., & Henriksen, R.N. 1993, *Astron. Astrophys.* **273**, 357

Raychaudhury, S. 1994, in *Cosmic Velocity Fields* (IAP, Paris), ed. F. Bouchet & M. Lachieze-Rey , in press

Regos, E., & Szalay, A.S. 1989, *Ap. J.* **345**, 627

Roth, J.R. 1994, in *Cosmic Velocity Fields* (IAP, Paris), ed. F. Bouchet & M. Lachieze-Rey , in press

Rowan-Robinson, M. 1993, *Proc. Nat. Acad. Sci.* **90**, 4822

Rowan-Robinson, M., Lawrence, A., Saunders, W., Crawford, J., Ellis, R.S., *et al.* 1990, *MNRAS* **247**, 1.

Rowan-Robinson, M., Lawrence, A., Saunders, W., & Leech, K. 1991, *MNRAS* **253**, 485

Rubin, V.C., Ford, W.K.Jr., Thonnard, N., Roberts, M.S., & Graham, J.A. 1976a, *Astron. J.* **81**, 687-718

Rubin, V.C., Thonnard, N., Ford, W.K.Jr., & Roberts, M.S. 1976b, *Astron. J.* **81**, 719-737

Rybicki, G.B., & Press, W.H. 1992, *Ap. J.* **398**, 169

Sachs, R.K., & Wolfe, A.M. 1967, *Ap. J.* **147**, 73

Sargent, W.L.W., & Turner, E.L. 1977, *Ap. J. Lett* **212**, L3

Scaramella, R., Vettolani, G., & Zamorani, G. 1991, *Ap. J. Lett* **376**, L1

Schechter, P. 1980, *Astron. J.* **85**, 801

Schuster, J. *et al.* 1993, *Ap. J. Lett* , in press

Seljak, U., & Bertschinger, E. 1993, *Ap. J.* , in press

Shaya, E., Peebles, P.J.E., & Tully, B. 1994, in *Cosmic Velocity Fields* (IAP, Paris), ed. F. Bouchet & M. Lachieze-Rey

Silk, J. 1989, *Ap. J. Lett* **345**, L11

Simmons, J.F.L., Newsam, A., & Hendry, M.A. 1994, in *Cosmic Velocity Fields* (IAP, Paris), ed. F. Bouchet & M. Lachieze-Rey , in press

Smoot, G.F., Gorenstein, M.V., & Muller, R.A. 1977, *Phys. Rev. Lett.* **39**, 898

Smoot, G.F. *et al.* 1992, *Ap. J. Lett* **396**, L1

Smoot, G.F., Tenorio, L., Banday, A.J., Kogut, A., Wright, E.L., *et al.* 1994, preprint

Stebbins, A. 1994, in *Cosmic Velocity Fields* (IAP, Paris), ed. F. Bouchet & M. Lachieze-Rey , in press

Strauss, M.A., Cen, R., & Ostriker, J.P. 1993, *Ap. J.* **408**, 389

Strauss, M.A., Cen, R., & Ostriker, J.P. 1994, in *Cosmic Velocity Fields* (IAP, Paris), ed. F. Bouchet & M. Lachieze-Rey , in press

Strauss, M.A., Davis, M., Yahil, A., & Huchra, J.P. 1990, *Ap. J.* **361**, 49

Strauss, M. A., Huchra, J. P., Davis, M., Yahil, A., Fisher, K. B., & Tonry, J 1992a, *Ap. J. Supp.* **83**, 29

Strauss, M. A., Yahil, A., Davis, M., Huchra, J. P., & Fisher, K. B. 1992b, *Ap. J.* **397**, 395

Szalay, A. 1988, in the Vatican Study Week on Large Scale Motions in the Universe, eds. G. V. Coyne & V. C. Rubin (Princeton: Princeton University Press) 323-38

Tonry, J.L. 1991, *Ap. J. Lett* **373**, L1

Tully, R.B., & Fisher, J.R. 1977, *Astron. Astrophys.* **54**, 661

Weinberg, D.H. 1991, *MNRAS* **254**, 315

Weinberg, D.H., & Cole, S. 1992, *MNRAS* **259**, 652

Willick, J. 1991, Ph.D. Thesis, UC Berkeley

Willick, J. 1994a, *Ap. J.* , in press

Willick, J. 1994b, *Ap. J.* , in press

Willick, J. *et al.* 1994, *Ap. J.* , in preparation

Yahil, A. 1994, *Ap. J. Lett* , submitted.

Yahil, A., Dekel, A., Kolatt, T., & Blumenthal, G. 1994, in preparation

Yahil, A., Walker, X., & Rowan-Robinson, M. 1986, *Ap. J. Lett* **301**, L1

Yahil, A., Strauss, M. A., Davis, M., & Huchra, J. P. 1991, *Ap. J.* **372**, 380

Zaroubi, S., & Hoffman, Y. 1994, preprint

Zel'dovich, Ya.B. 1970, *Astron. Astrophys.* **5**, 20